\documentclass[11pt, preprint]{aastex63}
\usepackage{rotating}
\usepackage{bm}
\graphicspath{{./}{figures/}}
\newcommand{\ii}{\mathrm{in}}
\newcommand{\oo}{\mathrm{out}}
\newcommand{\rpo}{r_\mathrm{p,\oo}}
\newcommand{\ain}{a_\ii}
\newcommand{\aout}{a_\oo}

\newcommand{\imut}{i_\mathrm{mut}}
\usepackage{xcolor}
\received{2022 July 18}
\revised{2022 September 1}
\accepted{2022 September 3}
\submitjournal{ApJ}
\shorttitle{Disruption timescales of hierarchical triples}
\shortauthors{Hayashi, Trani, Suto} 
\begin{document}

\title{Dynamical disruption timescales and chaotic behavior
    of hierarchical triple systems}
\correspondingauthor{Toshinori Hayashi}
\email{toshinori.hayashi@phys.s.u-tokyo.ac.jp}

\author[0000-0003-0288-6901]{Toshinori Hayashi}
\affiliation{Department of Physics, The University of Tokyo,  
Tokyo 113-0033, Japan}
\author[0000-0001-5371-3432]{Alessandro A. Trani}
\affiliation{Research Center for the Early Universe, School of Science,
The University of Tokyo, Tokyo 113-0033, Japan}
\affiliation{Okinawa Institute of Science and Technology Graduate University,
  Okinawa 904-0495, Japan}
\author[0000-0002-4858-7598]{Yasushi Suto}
\affiliation{Department of Physics, The University of Tokyo,  
Tokyo 113-0033, Japan}
\affiliation{Research Center for the Early Universe, School of Science,
The University of Tokyo, Tokyo 113-0033, Japan}
\affiliation{Laboratory of Physics, Kochi University of Technology,
  Tosa Yamada, Kochi 782-8502, Japan}

\begin{abstract}
	We examine the stability of hierarchical triple systems using
   direct $N$-body simulations without adopting a secular perturbation
   assumption. We estimate their disruption timescales in addition to
   the mere stable/unstable criterion, with particular attention to
   the mutual inclination between the inner and outer orbits. First,
   we improve the fit to the dynamical stability criterion by
   \citet{Mardling1999,Mardling2001} widely adopted in the previous
   literature. Especially, we find that that the stability boundary is
   very sensitive to the mutual inclination; coplanar retrograde
   triples and orthogonal triples are much more stable and unstable,
   respectively, than coplanar prograde triples.  Next, we estimate the
   disruption timescales of triples satisfying the stability condition
   up to $10^9$ times the inner orbital period. The timescales follow
   the scaling predicted by \citet{Mushkin2020}, especially at high
   $e_\oo$ where their random walk model is most valid.
   We obtain an improved empirical fit to the disruption timescales,
   which indicates that the coplanar retrograde triples are
   significantly more stable than the previous prediction. We
   furthermore find that the dependence on the mutual inclination can
   be explained by the energy transfer model based on a parabolic
   encounter approximation.  We also show that the disruption
   timescales of triples are highly sensitive to the tiny change of
   the initial parameters, reflecting the genuine chaotic nature of
   the dynamics of those systems.
 \end{abstract}


\section{Introduction \label{sec:intro}} 

Triple star systems ubiquitously exist in the
universe. Observationally, more than 70\% of OBA-type stars and 50\%
of FGK-type stars are known to belong to multiple systems
\citep[e.g.,][]{Raghavan2010,Sana2012}, and their multiplicity
increases with the mass of stars \citep{Moe2017}.  The diversity and
architecture of triple systems have been studied extensively through
recent photometric and spectroscopic observations. For example, 18
triples were recently resolved and their orbital elements determined
by combining proper motion and radial velocity
\citep[][]{Tokovinin2020, Tokovinin2021a}.  \citet{Hajdu2021}
discovered 23 triple systems using eclipse time variations with
OGLE-IV in eclipsing binaries. More recently, Gaia astrometric survey
reported the Data Release 3 (Gaia DR3), and about 100 genuine triple
stars were identified from astrometric and spectroscopic solutions
\citep{Gaia2022}. Gaia DR3 also reported 17 stars with potential
compact companions with large mass functions: $f(\mathrm{M})\equiv
m_2^3\sin^3{I}/(m_1+m_2)^2>3~M_\odot$. These candidates may include
binary compact objects, as well as single neutron star and black hole
companions.  These current observational results are quite
interesting, and suggest the importance of studying triple systems
both observationally and theoretically.

We are particularly interested in triple systems comprising an inner
binary black hole (BBH), which have not yet discovered.  While the
gravitational wave (GW) observations have revealed that there are
abundant populations of massive BBHs in the universe, their formation
mechanisms are not well understood; possible scenarios include the
evolution outcome of isolated massive binary stars
\citep[e.g.][]{Belczynski2002}, dynamical capture of two BHs in dense
stellar systems \citep[e.g.][]{Zwart2000}, and binary assembly from
primordial BHs \citep[e.g.][]{Sasaki2016}.  Regardless of the details
of such scenarios, the progenitors of the observed BBHs are likely to
spend a significantly long period as widely separated BBHs before the
coalescence.  Furthermore, a fraction of those progenitor BBHs may be
accompanied by a tertiary that accelerates the merger time through the
von Zeipel-Kozai-Lidov (ZKL) oscillations \citep{Blaes2002,Miller2002}.

More specifically, \citet{Antognini2016} performed $400$ million
simulations of gravitational scatterings of binary-binary,
triple-single, and triple-binary scatterings. They found that the
relatively close triples with a few - $100$ semi-major axis ratio is
efficiently produced from dynamical scattering events.  Later,
\citet{Fragione2020} performed systematic numerical simulations of
dynamical interactions of stars and compact bodies in dense clusters,
and found binary-binary encounters efficiently produced stellar and
compact triples. In addition, \citet{Trani2021} examines the formation
of wide-separate triples via dynamical interactions in young star
clusters from simulations performed in \citet{Rastello2020}, and found
that the ZKL oscillation plays an important role for the coalescence
of compact binaries in such triples. For more comprehensive reviews
on the formation of merging compact objects, including the
triple evolution channel, we refer to \citet{mapelli2021} and
\citet{spera_trani2022}.

In this context, we would like to mention an interesting triple system
comprising three compact objects, although not BHs \citep{Ransom2014}.
The triple was discovered by the pulsar timing, and consists of an
inner binary of a $1.4~M_\odot$ millisecond pulsar (PSR J0337+1715)
and $0.2~ M_\odot$ white dwarf, and a tertiary white dwarf of
$0.4~M_\odot$. Their inner and outer orbital periods are 1.63 days and
327 days, respectively, and the orbits are nearly coplanar
($i_\mathrm{mut} \approx 0.012~\mathrm{deg}$), and circular (the inner
eccentricity $e_\ii \approx 0.0007$ and outer eccentricity $e_\oo
\approx 0.04$). Even though its formation pathway would be very
  different from triples with inner BBHs, the above system proved the
  existence of triples consisting of compact objects.  Therefore, it
  is tempting to explore a feasibility to discovering an overlooked
  population of triples with such an amazing architecture.

These considerations motivated us to propose methodologies to search
for BBHs embedded in triple systems through the radial-velocity
modulations or the pulsar arrival time delays of tertiary
\citep{Hayashi2020,Hayashi2020_2,Hayashi2021}. They are expected to
provide a novel method to discover a population of BBHs in optical and
radio observations in a complementary fashion to the GW survey.  The
expected number of such triples with BBHs is very model-dependent and
difficult to predict theoretically. Thus a successful discovery of
such a triple would have a strong impact in understanding the
formation mechanism of BBHs as well.

Once such triples are formed, however, they may not be dynamically
stable and become disrupted in a relatively short timescale.  Indeed,
the orbital stability of triples is one of the long-standing key
questions in three-body problems, and there have been many previous
studies focusing on this issue.  Among others,
\citet{Mardling1999,Mardling2001} (hereafter, MA01) proposed a widely
used criterion of the dynamical instability of triples. Their
criterion was later generalized to include the dependence of inner
orbital elements \citep[e.g.][]{Myllari2018}, or general relativistic
corrections \citep[e.g.][]{Wei2021}. More recently,
  \citet{Lalande2022} proposed a machine learning method to predict
  the stability of the hierarchical triples.

In general, those criteria are intended to assess whether a triple is
stable or not, but cannot be directly used to predict their disruption
timescale. As we will show below, some systems remain bound even for
more than $10^8$ times the inner orbit period, depending on their
specific orbital configuration.  It is important, therefore, to find
an expression for the disruption timescale of triple systems.

\citet[][hereafter MK20]{Mushkin2020} have presented a physically
well-motivated model for that purpose. They particularly focus on a
triple system with a highly eccentric outer tertiary, and estimate the
disruption timescale by evaluating the energy transfer between the
inner and outer orbits based on the Random Walk (RW) assumption.  They
first consider a full RW model that numerically integrates the secular
evolution of triples, and propose an analytic expression for the
disruption timescale from a simplified version of the RW model.  Their
formula reasonably reproduces the numerical results of the disruption
time for triples whose outer eccentricity is sufficiently close to
unity, and thus the parabolic orbit approximation is justified, but
the agreement is significantly degraded otherwise.

The purpose of the present paper is to generalize the result of MK20
and find an empirical fit to the dynamical disruption timescales for
hierarchical triple systems over a larger parameter
space. Specifically, we proceed according to the following strategies.
\begin{enumerate}
\item We perform a series of systematic $N$-body simulations for
  hierarchical triple systems with various orbital configurations. We
  do not adopt the conventional orbit-average approximation, but
  directly simulate three point-mass particles using a fast and
  accurate $N$-body integrator \textsc{tsunami} \citep{trani2022b}.
\item As in MA01 and MK20, we consider Newtonian gravity alone, 
  neglecting general relativity or finite-size effects
  including spin and tidal interaction of particles.
\item We focus on the three specific values of the mutual inclination
  $\imut$: coplanar prograde ($\imut=0^\circ$), orthogonal
  ($\imut=90^\circ$), and coplanar retrograde ($\imut=180^\circ$).
\item We clarify the chaotic nature of the triple disruption by
  showing the high sensitivity to the input parameters, and
  thus attempt to obtain a statistical fit to the numerical results
  generalizing the RW model proposed by MK20.
\item We directly show that the scaling with respect to the mass and
  semi-major axis of triples is valid in a statistical manner despite
  their chaotic behavior of the disruption dynamics.
\item We apply the energy transfer model by \citet{Roy2003}
  to explain why coplanar retrograde triples are the most stable and
  initially orthogonal triples are the most unstable in general.
\end{enumerate}

The rest of the paper is organized as follows.  First, we briefly
review a few previous models relevant to the stability and disruption
timescales of hierarchical triples in section~\ref{sec:model}. Next,
we describe our numerical methods to integrate the triple systems
using the direct $N$-body simulation in section~\ref{sec:method}.
Section~\ref{sec:equalmass} focuses on triples comprising an
equal-mass inner binary, and discusses how sensitively the chaotic
nature of their dynamics changes the estimated timescales. In section
\ref{sec:comparison}, we show the stability boundaries and the
disruption timescales from our simulation runs, and obtain the
improved empirical fits by generalizing the previous models.
Section~\ref{sec:evolution} presents examples of evolution of orbital
elements, energy, and angular momentum of triple systems, and consider
how they depend on the mutual inclination between the inner and outer
orbits.  Finally, section~\ref{sec:summary} is devoted to summary and
conclusion of the paper. In appendix, we show how our estimates of
disruption timescales are affected by the variation of initial phase
angles (section \ref{sec:Td-phase}), summarize the ejection pattern
(section \ref{sec:ejection}), and discuss the possible effect of
general relativity that is ignored in the present paper
(section~\ref{sec:grprecession}).

\section{Previous models for the dynamical stability of triple systems
  \label{sec:model}}

\begin{figure*}
\begin{center}
\includegraphics[clip,width=10cm]{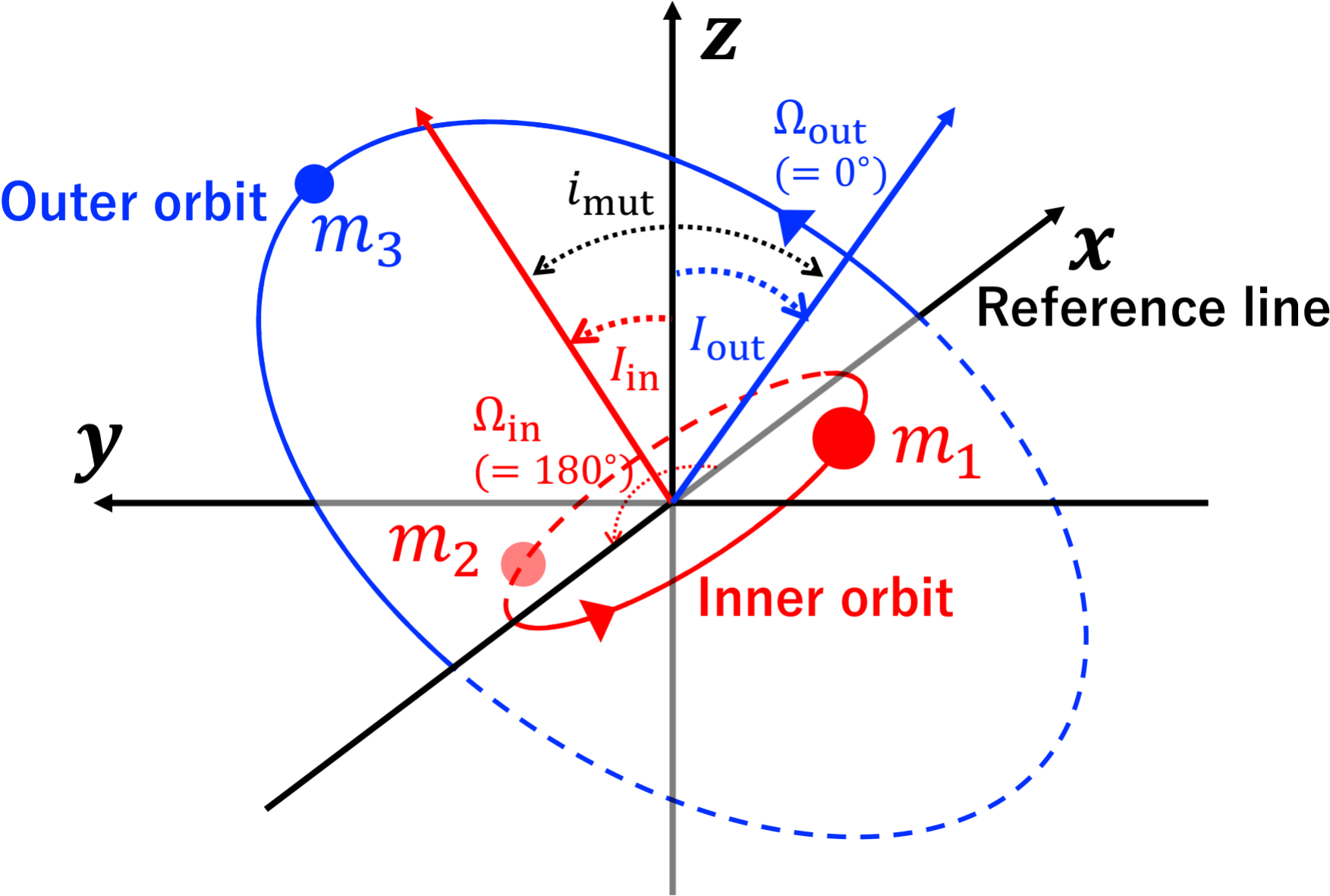}
\end{center}
\caption{Orbital configuration of a triple system consisting of an
  inner binary and a tertiary. The origin of the reference frame is
  set to be the barycenter of the inner binary, instead of the triple.
  \label{fig:schematic_triple}}  
\end{figure*}

Figure \ref{fig:schematic_triple} illustrates a schematic
configuration of a hierarchical triple system considered in the
present paper.  The inner binary consists of two massive bodies of
$m_1$ and $m_2$, and a tertiary with mass $m_3$ orbits around the
binary.  The orbital parameters of inner and outer orbits are
specified in terms of the Jacobi coordinates. Each orbit is
characterized by the semi-major axis $a_j$, the eccentricity $e_j$,
and the argument of pericenter $\omega_j$. The orientation of orbits
in space are specified by the longitude of ascending nodes $\Omega_j$
measured from the reference line, and the mutual inclination
$i_\mathrm{mut}$ between inner and outer orbits.

This section briefly reviews three important previous models that
underlie our current approach to estimating the dynamical disruption
time of triple systems.

\subsection{Dynamical instability criterion by MA01}

\citet{Mardling1995,Mardling1995b,Mardling1999,Mardling2001}
considered evolution of a tidally interacting binary.  They defined
that a binary evolution is chaotic when the direction and amount of
energy flows between the tide and orbit become entirely unpredictable.
In reality, the boundary between normal and chaotic behavior is not
well defined, since it is extremely sensitive to small differences of
orbital parameters and regions of stability and chaos are intertwined
in a complicated fashion.  Nevertheless, they were able to derive an
approximate formula for the boundary.

Remarkably, MA01 recognized that the way of energy and angular
momentum exchange between the inner and outer orbits in a hierarchical
triple is very similar to that in a tidally interacting binary.  On
the basis of the analogy, MA01 proposed the following dynamical
instability criterion for hierarchical triples:
\begin{eqnarray}
\label{eq:MAcriterion}
\frac{\rpo}{\ain} \equiv \frac{\aout(1-e_\oo)}{\ain}
< \left(\frac{\rpo}{\ain}\right)_\mathrm{MA}
  \equiv 2.8\left(1-0.3\frac{i_\mathrm{mut}}{\pi}\right)
  \left[\left(1+\frac{m_3}{m_{12}}\right)
    \frac{(1+e_\oo)}{\sqrt{1-e_\oo}}\right]^{2/5} .
\end{eqnarray}
In the above expression, $\ain$ and $\aout$ are the semi-major axes of
the inner and outer orbits, $e_\oo$ is the eccentricity of the outer
orbit, $i_\mathrm{mut}$ is the mutual inclination between the two
orbits, and $m_{12}=m_1+m_2$ is the total mass of the inner binary.

The numerical coefficient of 2.8 in inequality~(\ref{eq:MAcriterion})
is empirically introduced by MA01 from numerical simulations. They
define the stability based on whether two orbits, differing by $1$
part in $10^5$ in the eccentricity, remain similar after 100 outer
orbits.  The additional dependence on the mutual inclination,
$(1-0.3i_\mathrm{mut}/\pi)$, is heuristically introduced in
\citet{Aarseth2001} based on the result of stability enhancement due
to $i_\mathrm{mut}$ discussed by \citet{Harrington1972}. The extension
of this criterion is done by \citet{Myllari2018} to include the
dependence on the mass ratio of the inner binary $m_{2}/m_{1}$, the
mutual inclination $i_\mathrm{mut}$, and the inner eccentricity
$e_\ii$.

Inequality~(\ref{eq:MAcriterion}) indicates that the ratio between the
pericenter distance of the outer orbit and the semi-major axis of the
inner orbit is the key parameter dictating the chaotic behavior of the
triples. Triples with $\rpo/\ain <1$ are unstable because their outer
and inner orbits cross each other.  Inequality~(\ref{eq:MAcriterion})
should be interpreted as a criterion for the onset of chaotic
evolution, rather than for disruption, but triples with $1<\rpo/\ain
<(\rpo/\ain)_\mathrm{MA}$ begin chaotic orbital evolution and are
expected to be disrupted eventually.  Therefore, while inequality
(\ref{eq:MAcriterion}) is widely known as a standard criterion for
dynamically unstable triples
\citep[e.g.][]{Perpinya2019,Gupta2020,Tanikawa2020}, it cannot be used
to predict the disruption time.

\subsection{Random Walk Model by  MK20}

In order to estimate the disruption time of triples, MK20 apply the
random walk (RW) approximation for the energy transfer between the
inner and outer orbits.  They focus on very eccentric outer orbits
($e_\oo \sim 1$), and use an approximate analytic formula of the
energy during the pericenter passages of a parabolic orbit by
\citet{Roy2003}. The transferred energy depends on the inner mean
anomaly during the next outer orbital pericenter. Their RW model
assumes that the inner mean anomaly changes randomly between adjacent
outer pericenter passages without net energy exchange on average. The
resulting energy transfer process, therefore, is supposed to be well
described by a random walk.  Furthermore, they assume that the
disruption likely occurs when the accumulated transferred energy
variance becomes comparable to the square of the outer orbital energy,
and derive the disruption time of triples.

In their full RW model, they perform secular integration numerically
up to the quadrupole interaction so as to include the effect of
evolving orbital parameters. They also consider a simplified version
of the RW model, in which the complicated evolution of orbital
elements are neglected. Their simplified RW model yields the following
analytic expression for the disruption time $T_\mathrm{d}$:
\begin{eqnarray}
\label{eq:rwmodel}
\frac{T_\mathrm{d}}{P_\ii}
= 2\left(\frac{m_{123}m_{12}}{m_1 m_2}\right)^2\sqrt{1-e_\oo}
\left(\frac{\rpo}{\ain}\right)^{-2}
\exp\left[\frac{4\sqrt{2}}{3}\sqrt{\frac{m_{12}}{m_{123}}}
\left(\frac{\rpo}{\ain}\right)^{3/2}\right],
\end{eqnarray}
where $P_\ii$ is the initial orbital period of the inner binary, and
$m_{123}\equiv m_1+m_2 +m_3$ is the total mass of the system.

The factor $2$ in equation (\ref{eq:rwmodel}) is heuristically
determined from their $N$-body simulations. They found that the RW
model estimations can reasonably reproduce the disruption times of
numerical simulations within one order-of-magnitude for high-$e_\oo$
triples; they also perform simulations for $e_\oo=0.1$, 0.3, 0.5, 0.7,
and $0.9$, and confirm that their models agree well with the
simulations down to $e_\oo = 0.7$ although the accuracy of equation
(\ref{eq:rwmodel}) significantly decreases at lower eccentricities.

\subsection{Energy transfer for a parabolic encounter
  \label{subsec:roy2003}}

The energy transfer model adopted by MK20 is based on a binary-single
scattering process for a parabolic encounter by
\citet{Roy2003}. Indeed, this model turns out to be important to
understand the dependence of the disruption time on the mutual
inclination, we briefly summarize the result of \citet{Roy2003} in
this subsection.

They apply the perturbation theory and compute the energy exchange
during an encounter in which the collision parameter is much larger
than the semi-major axis of a binary, neglecting rapidly oscillating
terms.  If the eccentricity of outer orbit is close to unity, the
energy transfer during a pericenter passage is well approximated as a
binary-single scattering for a parabolic encounter.  The energy
transfer during one pericenter passage for a triple system is written
as \citep[][]{Mushkin2020}:
\begin{eqnarray}
  \label{eq:dE-W-F}
  \Delta E_\oo = W(a_\ii,r_\mathrm{p,\oo},m_1,m_2,m_3)
  F(\phi,\Omega,i_\mathrm{mut},e_\ii),
\end{eqnarray}
where $r_\mathrm{p,\oo}$, $\phi$, and $\Omega$ are the pericenter
distance of the outer orbit, the characteristic phase angle, and the
relative longitude of the ascending nodes, respectively.

The characteristic phase $\phi$ is defined as
\begin{eqnarray}
	\phi \equiv 2\omega - M^{*}_\ii, 
\end{eqnarray}
where $\omega$ and $M^{*}_\ii$ are the pericentre argument, and the
inner mean anomaly during the passage, respectively. In equation
(\ref{eq:dE-W-F}),
\begin{eqnarray}
\label{eq:def-W}
  W \equiv -E_\ii \frac{m_3}{m_{12}}
  \left(\frac{m_{12}}{m_{123}}\right)^{5/4}
  \left(\frac{r_\mathrm{p,\oo}}{a_\ii}\right)^{3/4}
  \exp\left[-\frac{4m_{12}}{3m_{123}}
\left(\frac{r_\mathrm{p,\oo}}{a_\ii}\right)^{3/2}
\right],
\end{eqnarray}
where
\begin{eqnarray}
	E_\ii = -\frac{\mathcal{G}m_1m_2}{2a_\ii}.
\end{eqnarray}

The function $F$ in equation (\ref{eq:dE-W-F}) is given by
\begin{eqnarray}
\label{eq:F-A}
  F \equiv \sqrt{2}A_1\sin{\phi}+ 
2A_2\sin{\phi}\cos{2\Omega}
+ 2A_3\cos{\phi}\sin{2\Omega}, 
\end{eqnarray}
and the dependence on $e_\ii$ and $i_\mathrm{mut}$ is explicitly
included in the three coefficients:
\begin{eqnarray}\label{eq:A1}
  \label{eq:def-A1}
  A_1(e_\ii,i_\mathrm{mut}) &=&
  \frac{\sqrt{\pi}(f_2-f_1)}{2^{1/4}}\sin^2{i_\mathrm{mut}},
  \\\label{eq:A2}
  A_2(e_\ii,i_\mathrm{mut}) &=&
  -2^{5/4}\sqrt{\pi}f_4\cos{i_\mathrm{mut}}
  -\frac{\sqrt{\pi}(f_1+f_2)}{2^{7/4}}(3+\cos{2i_\mathrm{mut}}),
  \\ \label{eq:A3}
  A_3(e_\ii,i_\mathrm{mut}) &=& -2^{1/4}\sqrt{\pi}(f_1+f_2)\cos{i_\mathrm{mut}}
  -\frac{\sqrt{\pi}f_4}{2^{3/4}}(3+\cos{2i_\mathrm{mut}}). 	
\end{eqnarray}
In the above expressions, $f_i$ are written in terms of
the Bessel function of the $i$-th order $J_{i}$:
\begin{eqnarray}
f_1 &=& J_{-1}(e_\ii)-2e_\ii J_{0}(e_\ii) + 2e_\ii J_2(e_\ii)
  - J_3(e_\ii),  \\
f_2 &=& (1-e_\ii^2)[J_{-1}(e_\ii)-J_{3}(e_\ii)],  \\ 
\label{eq:def-f4}
f_4 &=& \sqrt{1-e_\ii^2}[ J_{-1}(e_\ii)-e_\ii J_{0}(e_\ii)
          -e_\ii J_2(e_\ii) + J_3(e_\ii)]. 
\end{eqnarray}

In section \ref{sec:evolution}, we show that the energy transfer is
suppressed for retrograde orbits using these expressions, and discuss
a possible origin of the dependence of orbital configurations on the
disruption timescales.

\section{Numerical methods}\label{sec:method}

\subsection{Simulation parameters and scaling}

As illustrated in Figure \ref{fig:schematic_triple}, there are a
number of parameters characterizing triples, and it is not realistic
to cover the parameter space systematically.  Thus, we adopt a
fiducial set of parameters, which is listed in Table 1. The specific
choice of $m_{12}$ is partly motivated by our previous papers
\citep{Hayashi2020,Hayashi2020_2} that consider a triple comprising an
inner BBH and a stellar tertiary. Due to the scaling in the basic
equations, however, the result can be easily applied for other
cases. More specifically, mass and length, and time scales in purely
gravitational problem admit the following scaling relation:
\begin{equation}
\label{eq:scaling}
  M'=\alpha M, \quad L'=\beta L,
  \quad T'=\gamma T= \sqrt{\frac{\beta^3}{\alpha}}T.
\end{equation}
The relation (\ref{eq:scaling}) implies that the disruption time
$T_\mathrm{d}$ computed from our set of simulation runs can be scaled
as $T'_\mathrm{d}=\sqrt{\beta^3/\alpha}~T_\mathrm{d}$ Indeed, this is
why we mostly express our results below in terms of dimensionless
variables like $T_\mathrm{d}/P_\ii$.

Therefore, more relevant parameters are the mass ratio of the inner
binary $q_{21}=m_2/m_1$, the mass ratio of secondary and tertiary
$q_{23}=m_2/m_3$, the outer pericenter distance over inner semi-major
axis $\rpo/a_\ii$, the outer eccentricity $e_\oo$ and the mutual
inclination $i_\mathrm{mut}$.

In addition to those parameters, the variation of initial phase angles
($\omega_\ii$, $\omega_\oo$, $M_\ii$, and $M_\oo$) would introduce
statistical variance on $T_\mathrm{d}/P_\ii$. As will be discussed in
appendix, we find that the resulting stochastic variance amounts to
one or two orders of magnitudes, which is indeed comparable to the
level of fluctuations due to the intrinsic chaotic nature of the
triple systems that we consider in this paper.

\begin{deluxetable}{lcccccccccc}
\tablecolumns{3}
\tablewidth{1.0\columnwidth} 
\tablecaption{Fiducial values for the simulated triple systems} 
\tablehead{ orbital parameter& symbol  & initial value}
\startdata
inner-binary mass  & $m_{12}$ & $20~M_\odot$ \\
inner-binary period & $P_\mathrm{in}$ & $1000~\mathrm{days}$ \\
inner eccentricity & $e_\mathrm{in}$ & $10^{-5}$ \\
inner pericenter argument & $\omega_\mathrm{in}$ & $180~\mathrm{deg}$ \\
outer pericenter argument & $\omega_\mathrm{out}$ & $0~\mathrm{deg}$ \\
inner longitude of ascending node & $\Omega_\ii$ & $180~\mathrm{deg}$ \\
outer longitude of ascending node & $\Omega_\oo$ & $0~\mathrm{deg}$ \\
inner mean anomaly & $M_\ii$ & $30~\mathrm{deg}$ \\
outer mean anomaly & $M_\oo$ & $45~\mathrm{deg}$ \\ 
\hline
\hline
\enddata
\label{tab:fiducial}
\tablecomments{We adopt the above values for input of simulations as
  our fiducial systems unless otherwise specified. We vary the other
  parameters including $i_\mathrm{mut}$, $q_{21}\equiv m_2/m_1(\le
  1)$, $q_{23}\equiv m_2/m_3(>0)$, $e_\oo$, and $\rpo/a_\ii$ as
  discussed in the main text. Note that $a_\ii$ is uniquely determined
  by $m_{12}$ and $P_\ii$, and $m_1$, $m_2$ and $m_3$ are uniquely
  determined by $m_{12}$, $q_{21}$, and $q_{23}$.}
\end{deluxetable}

\begin{figure*}
\begin{center}
\includegraphics[clip,width=7cm]{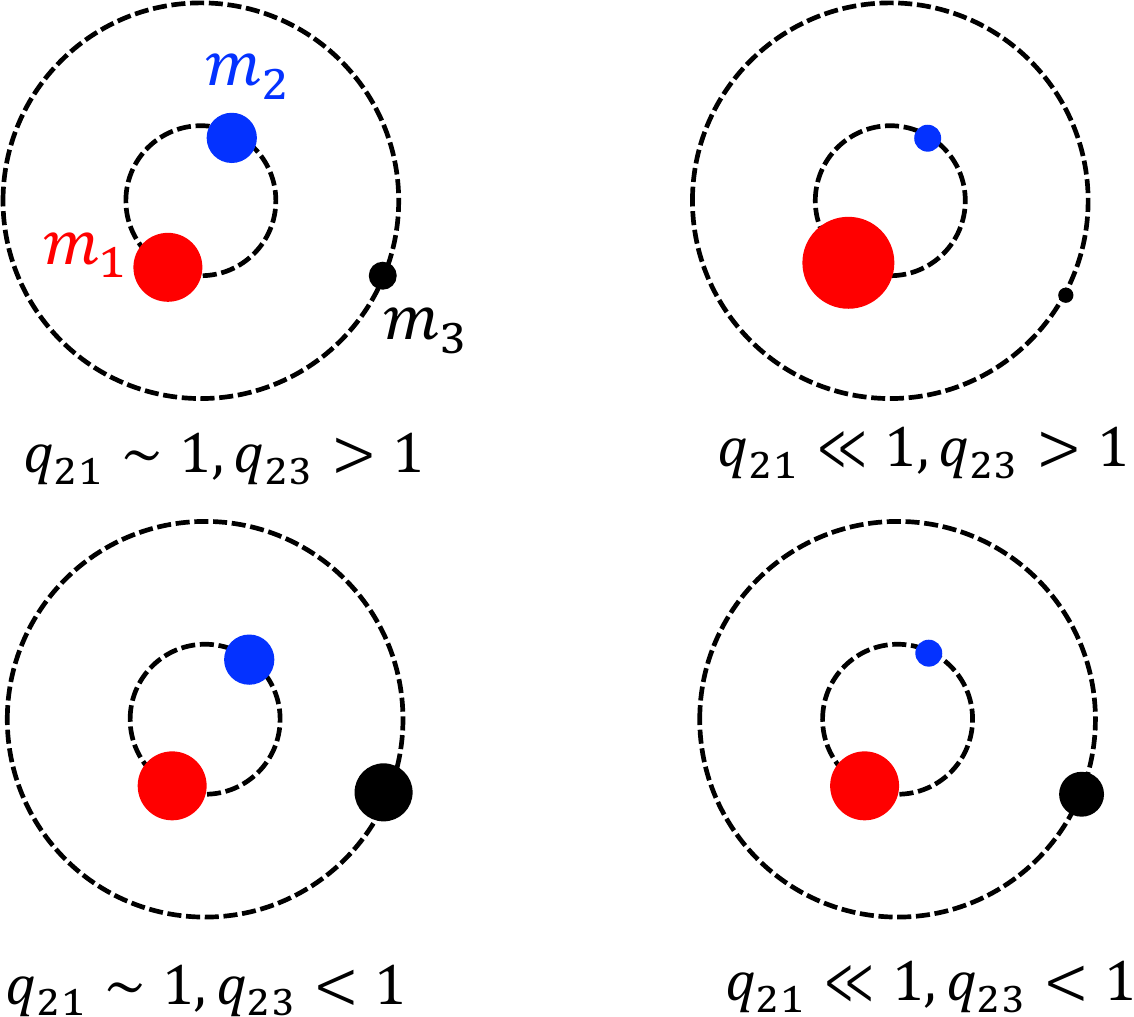}
\end{center}
\caption{ Schematic illustrations of triple systems for very different
  $q_{21}$ and $q_{23}$ regimes. Red, blue, and black filled circles
  correspond to the primary, secondary, and tertiary, respectively.
  Note that a triple with $q_{23}<1$ indicates that
    its tertiary is more  massive than the secondary.
	\label{fig:4orbits}}  
\end{figure*}

We define the primary as the most massive body in the inner binary (so
that $m_1 \geq m_2$), and thus $0 < q_{21} \leq 1$.  While $q_{23}>1$
for a hierarchical triple consisting of inner massive binary and a
tertiary, we consider cases of $q_{23}<1$ as well that are relevant
for planetary systems in general.  Figure \ref{fig:4orbits} illustrates
schematic pictures of triples in different $q_{21}$ and
$q_{23}$ regimes.

\subsection{\texorpdfstring{$N$}{N}-body integrator}

In order to obtain the disruption time distributions of triples, we
perform a series of numerical simulations with the $N$-body integrator
{\tt TSUNAMI} \citep[][Trani et al., in prep.]{Trani2019}.  {\tt
  TSUNAMI} is a direct $N$-body integrator specifically designed to
accurately simulate few-body systems. The code solves the Newtonian
equations of motion derived from a time-transformed, extended
Hamiltonian \citep{Mikkola2013a,Mikkola2013b}. This numerical
procedure, also called algorithmic regularization, serves to avoid the
well known issue of gravitational integrators when two particles get
very close together. When no regularization is employed, the
acceleration grows quickly, and the timestep becomes extremely small,
sometimes even halting the integration. Specifically, we employ
  the logarithmic Hamiltonian regularization scheme, which corresponds
  to setting $\alpha = 1$, $\beta = 0$, and $\gamma = 0$ in the
  notation of \citet{Aarseth2008}.  Furthermore, {\tt TSUNAMI} is
suited to evolve hierarchical systems of particles like the ones
examined here, thanks to its chain-coordinate system that reduces
round-off errors when calculating distances between close particles
far from the center of mass of the system
\citep{Mikkola1993}. Finally, {\tt TSUNAMI} uses Bulirsch-Stoer
extrapolation to improve the accuracy of the integration, ensuring
accuracy and adaptability over a wide range of dynamical scales
\citep{BulirschStoer2002}. As a result, {\tt TSUNAMI} can perform
accurate three-body simulations roughly $10 - 100$ times faster than
other integrators. For more details on the integration scheme, we
refer to \citet{trani2022b}.

The definition of the dynamical stability is not unique. Indeed, it
may depend on the maximum integration time that we adopt in numerical
simulations to some extent. First, we define the disruption time
$T_\mathrm{d}$ for each triple following
\citet{Manwadkar2020,Manwadkar2021}. At each timestep, the integrator
evaluates the binding energy for each of the three pairs of bodies,
{\it i.e.} $(m_1, m_2)$, $(m_1,m_3)$ and $(m_2,m_3)$. The pair with
the highest (negative) binding energy is considered as the inner
binary, and we call the pair of the inner binary and the remaining
tertiary as the outer pair. When the binding energy of the outer pair
becomes positive and the radial velocity of the tertiary becomes
positive (i.e. is moving away from the inner binary), we
  tentatively regard it as a candidate of a disrupted triple, and
  assign the epoch as the disruption time $T_\mathrm{d}$.  Some of
  them, however, represent a transient unbound state, and subsequently
  become gravitationally bound again. Therefore, we continue the run
  until its binary-single distance exceeds 20 times the binary
  semi-major axis so as to make sure that it is a truly disrupted
  triple. In that case, we stop the run, and the triple is classified
  as disrupted with the previously assigned value of $T_\mathrm{d}$.
  If a system becomes bound again, on the other hand, we
  reset the temporarily assigned value of $T_\mathrm{d}$,
  and keep running the integration of the system.

Finally, we define {\it stable} triples as those that do not
  disrupt and survive until our maximum integration time
  $t_\mathrm{int}$.  We set $t_\mathrm{int}= 10^{9} P_\ii$ unless
otherwise specified.

Due to the chaotic nature of the system, strict numerical
  convergence cannot be achieved, and the disruption timescales
  obtained from our simulation runs should be interpreted in a
  statistical manner. In \S \ref{subsec:chaos-scaling} and
  appendix~\ref{sec:Td-phase}, we attempt a numerical convergence test
  by running the same set of runs on different machines allowing for a
  tiny difference of the initial orbital parameter, as well as varying
  the initial phases of the inner and outer orbits and of the three
  bodies. In all cases, the disruption timescales agree within one or
  two orders of magnitudes, indicating that our results are
  statistically robust within those uncertainties.

\section{Qualitative behavior of hierarchical triples
  comprising an equal-mass inner binary
  \label{sec:equalmass}}

\subsection{Disruption timescales estimated from simulations}
\label{subsec:equalmass-sim}

This subsection presents our simulation results of the disruption
timescales, focusing on triples with an equal-mass inner binary.  More
general cases will be described in section \ref{sec:comparison}.

\begin{figure*}
\begin{center}
\includegraphics[clip,width=9.0cm]{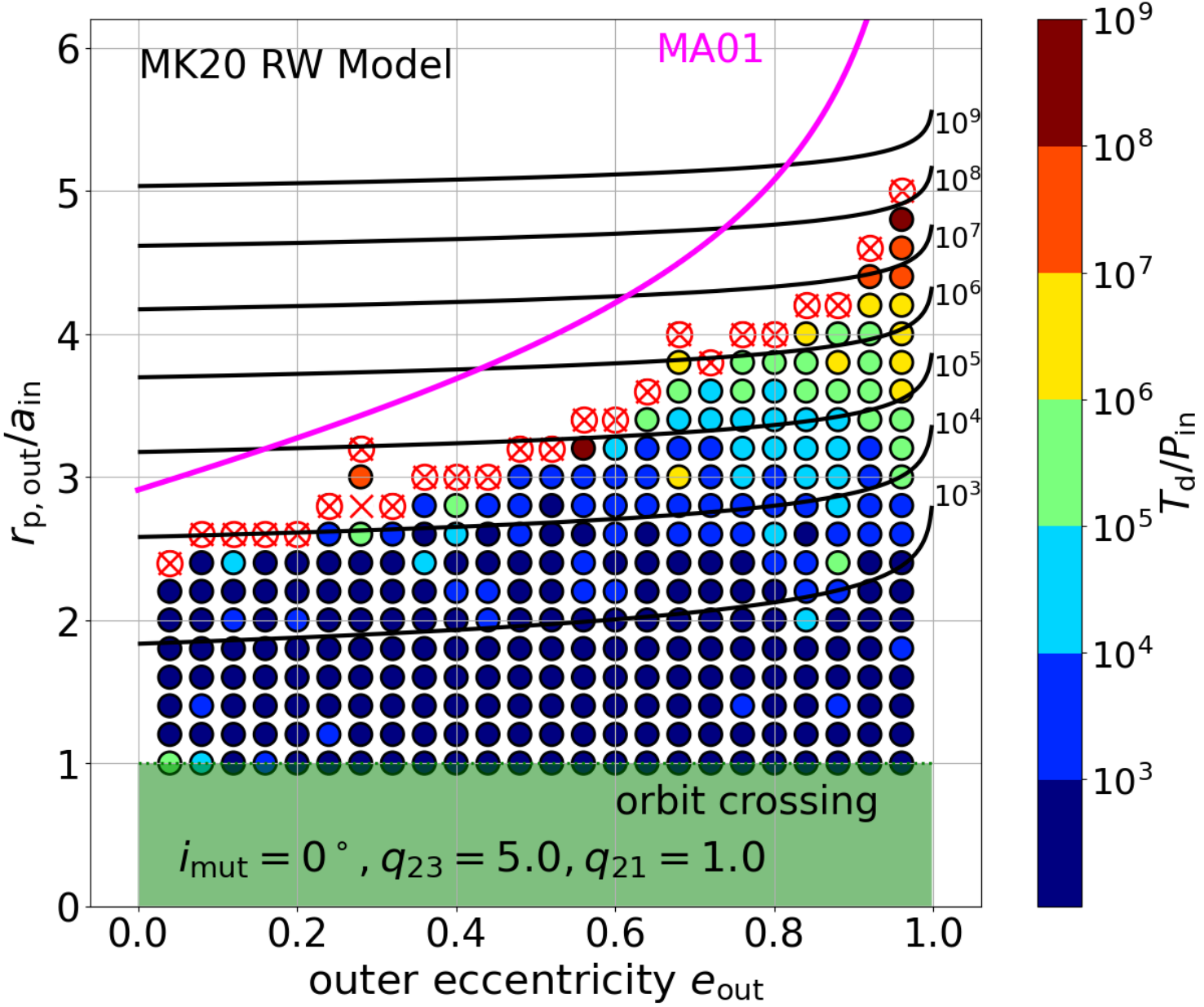}
\end{center}
\caption{The disruption time $T_\mathrm{d}/P_\ii$ dependence on $e_\oo
  - \rpo/a_\ii$ plane for the coplanar prograde ($\imut=0^\circ$) and
  equal-mass ($q_{21}=1$) inner binary case with $q_{23}=5$. The
  disruption timescales evaluated from simulations are indicated
  according to the side color scales. Magenta and black curves
  represent the MA01 dynamical stability criterion and the MK20 model
  predictions, respectively. The region where the inner and outer
  orbits of the system cross ($\rpo/a_\ii<1$) is shaded in
  green. Cross symbols indicate the triples that remain stable until
  $t_\mathrm{int}=10^9 P_\ii$, while circle-crosses symbols define the
  boundary of triples that disrupt before $10^9 P_\ii$.
\label{fig:dist_0}}
\end{figure*}

Figure \ref{fig:dist_0} shows the normalized disruption time
$T_\mathrm{d}/P_\ii$ distribution on $e_\oo$ - $\rpo/a_\ii$ plane for
coplanar prograde triples with an equal-mass inner binary
($i_\mathrm{mut}=0^\circ, q_{21}=1$). The secondary to tertiary mass
ratio $q_{23}$ is fixed as $5.0$, {\it i.e.,} the tertiary mass $m_3$
is $2M_\odot$ for $m_1=m_2=10M_\odot$. We adopt the fiducial values
for the other parameters that are listed in Table \ref{tab:fiducial}.
As discussed in section \ref{subsec:chaos-scaling}, $T_\mathrm{d}$
does not exactly scales with $P_\ii$ because of the scatters
inevitably induced by the chaoticity.  Nevertheless,
$T_\mathrm{d}/P_\ii$ is independent of $P_\ii$ statistically, and we
consider the normalized disruption time $T_\mathrm{d}/P_\ii$ in what
follows.

Triples located above the magenta curves in Figure \ref{fig:dist_0}
are expected to be stable according to the MA01 dynamical stability
condition.  The disruption timescales predicted by the MK20 model are
plotted in black contours (for $T_\mathrm{d}/P_\ii=10^3, 10^4, 10^5,
10^6, 10^7, 10^8$ and $10^9$). The contours are nearly straight lines
on $e_\oo$ - $\rpo/a_\ii$ plane because the exponential term in
equation (\ref{eq:rwmodel}) dominates the parameter dependence.  The
green shaded region in Figure \ref{fig:dist_0} corresponds to the
orbit-crossing triples whose outer orbit intersects the inner binary
orbit at the pericentre, $a_\oo(1-e_\oo)<a_\ii$. Those triples are
supposed to be very unstable in general, and are not our main
interest in the present paper.

The result of simulated disruption timescales are plotted as filled
circles with different colors in Figure \ref{fig:dist_0} according to
the side color-scale. We start the simulation with the initial
parameters of $(\rpo/a_\ii,e_\oo)=(1.0,0.04+0.04m)$ with the integer
$m$ running over $0-23$. For a given value of $m$, we sequentially
increase the value of $\rpo/a_\ii$ by 0.2.  Due to the limitation of
the CPU time, we stop each simulation at $t_\mathrm{int}=1.0\times10^9
P_\ii$ even if the system is not disrupted, which is approximately
equal to $2.7$ Gyrs for our fiducial parameters. For any given $m$, we
stop increasing the value of $\rpo/a_\ii$ when the previous two
realizations at lower $\rpo/a_\ii$ do not break before the final
integration time. In other words, we scan the $\rpo/a_\ii$ - $e_\oo$
of Figure~\ref{fig:dist_0} going from bottom to top, stopping only
when two realizations in a row fail to break up before
$t_\mathrm{int}$. Thus those triples located above the circle-cross
symbols in Figure \ref{fig:dist_0} correspond to
$T_\mathrm{d}>1.0\times 10^9P_\ii$. In what follows, we sometimes
refer to the boundary as the simulated stability boundary just for
convenience.
 
Figure \ref{fig:dist_0} shows that the MA01 dynamical stability
condition, inequality (\ref{eq:MAcriterion}), roughly explains the
stability boundary of the simulated coplanar prograde triples with an
equal-mass inner binary; all triples with $\rpo/\ain >
({\rpo}/{\ain})_\mathrm{MA}$ are stable at least until
$t_\mathrm{int}$.
 
The predicted disruption timescales based on the MK20 RW model agree
well with the simulation result only for triples with $e_\oo \gtrsim
0.8$.  This is indeed expected because their model assumes that the
energy transfer between the inner and outer orbits is approximated by
a random walk for a highly eccentric outer orbit. MK20 also confirmed
that their RW model reproduces their simulations reasonably well down
to $e_\oo \sim 0.7$, which is consistent with our present result. The
right panel in Figure \ref{fig:dist_0}, furthermore, implies that the
lower limit of $e_\oo$ where the RW model approximation is valid
becomes larger as $\rpo/a_\ii$ increases.

\begin{figure*}
\begin{center}
\includegraphics[clip,width=9.0cm]{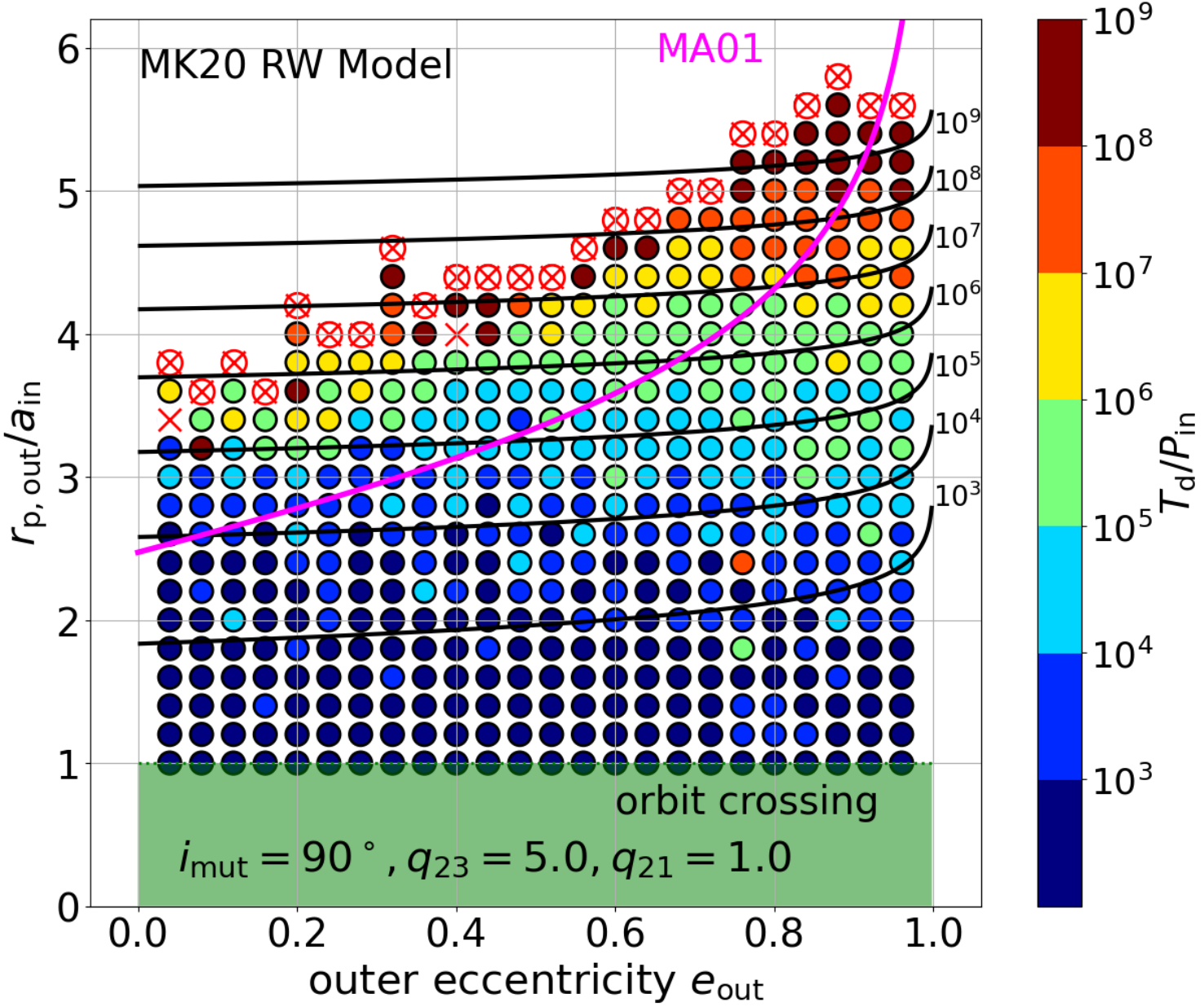}
\end{center}
\caption{Same as Figure \ref{fig:dist_0} but for initially orthogonal
  triples ($i_\mathrm{mut}=90^\circ$).\label{fig:dist_90}}
\end{figure*}

\begin{figure*}
\begin{center}
\includegraphics[clip,width=9.0cm]{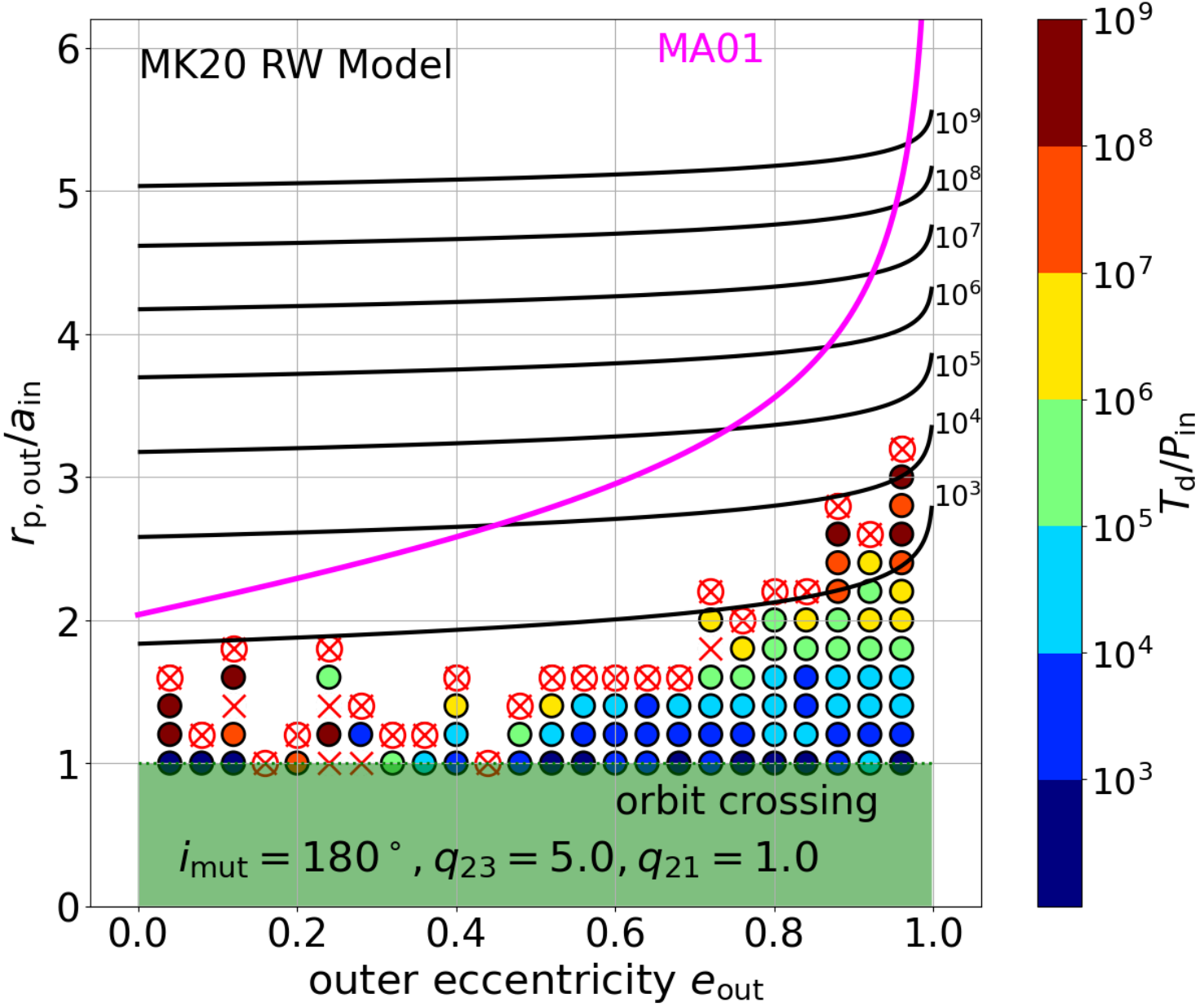}
\end{center}
\caption{Same as Figure \ref{fig:dist_0} but for coplanar retrograde
  triples ($i_\mathrm{mut}=180^\circ$). \label{fig:dist_180}}
\end{figure*}

In reality, however, the above rough agreement between the
  theoretical predictions and simulations does not hold for orthogonal
  ($i_\mathrm{mut}=90^\circ$) and coplanar retrograde
  ($i_\mathrm{mut}=180^\circ$) cases; see Figures \ref{fig:dist_90}
  and \ref{fig:dist_180}.

Orthogonal triples are less stable than the MA01 criterion,
  which may be due to their drastic orbital evolution. In particular,
  the effect of the ZKL oscillation is not properly taken into account
  in the MA01 model. In contrast, the MK20 RW model seems to agree
  better than the prograde simulations for even lower
  values of $e_\oo$. Finally, Figure \ref{fig:dist_180} clearly
  indicates that coplanar retrograde triples are significantly more
  stable than MA01 and MK20 predict.  We will discuss this point later
  based on the energy transfer model by \citet{Roy2003}.

\subsection{Chaotic nature of triple dynamics and the scaling relation}
\label{subsec:chaos-scaling}

The results in Figures \ref{fig:dist_0} to \ref{fig:dist_180} are
computed initially for a set of fixed phase parameters including mean
anomalies ($M_\ii$, $M_\oo$) and arguments of pericenter
($\omega_\ii$, $\omega_\oo$).  While all those initial parameters are
mixed up in the course of the dynamical evolution, their different
values at the initial epoch are likely to affect the disruption
timescales to some extent, and to introduce statistical scatters in
the estimated timescales. This is quite expected, and we show in
appendix \ref{sec:Td-phase} that those phase angles change the
disruption timescales by one or two orders of magnitudes.

\begin{figure*}
\begin{center}
\includegraphics[clip,width=8cm]{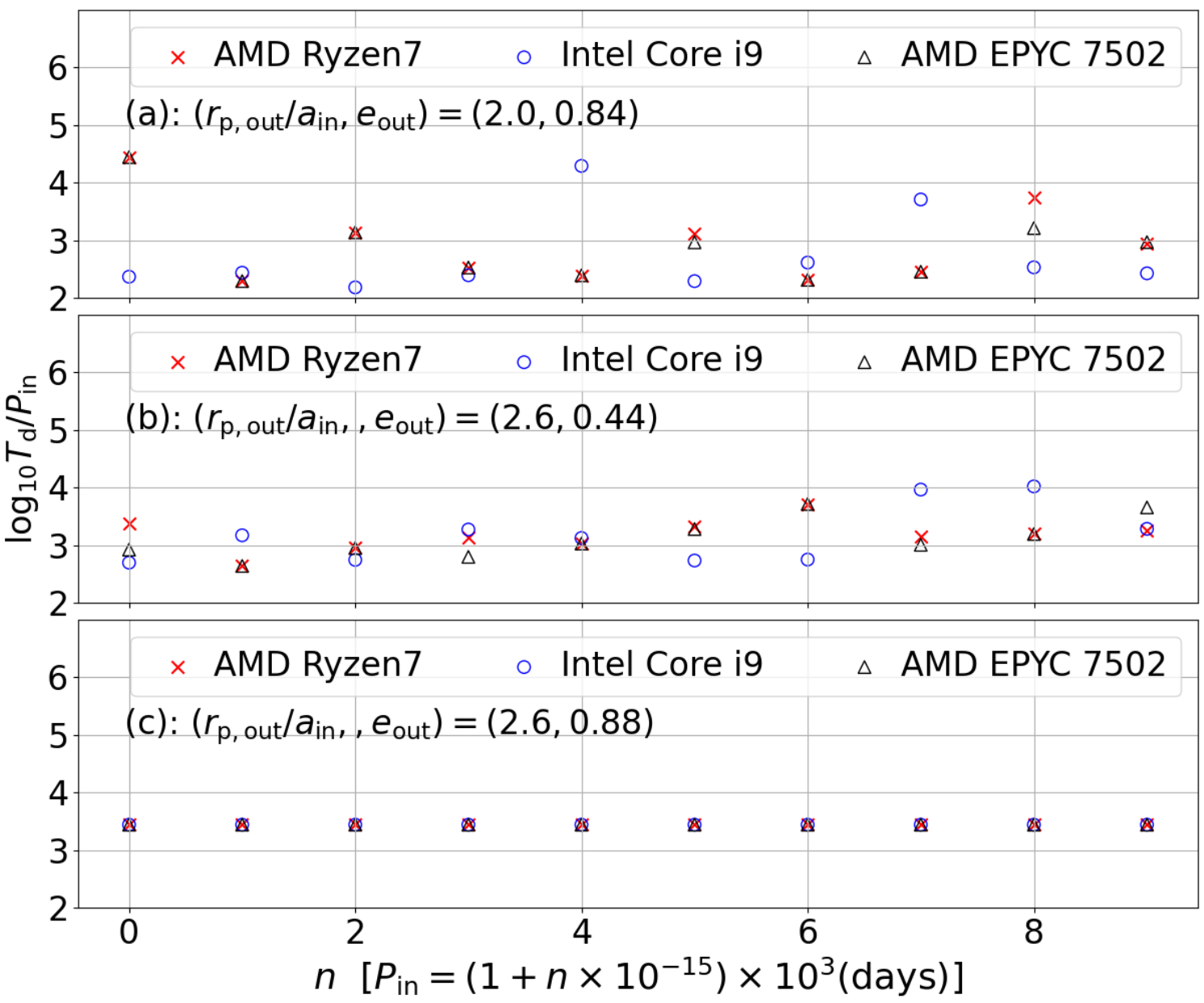}
\hspace{5pt}
\includegraphics[clip,width=8cm]{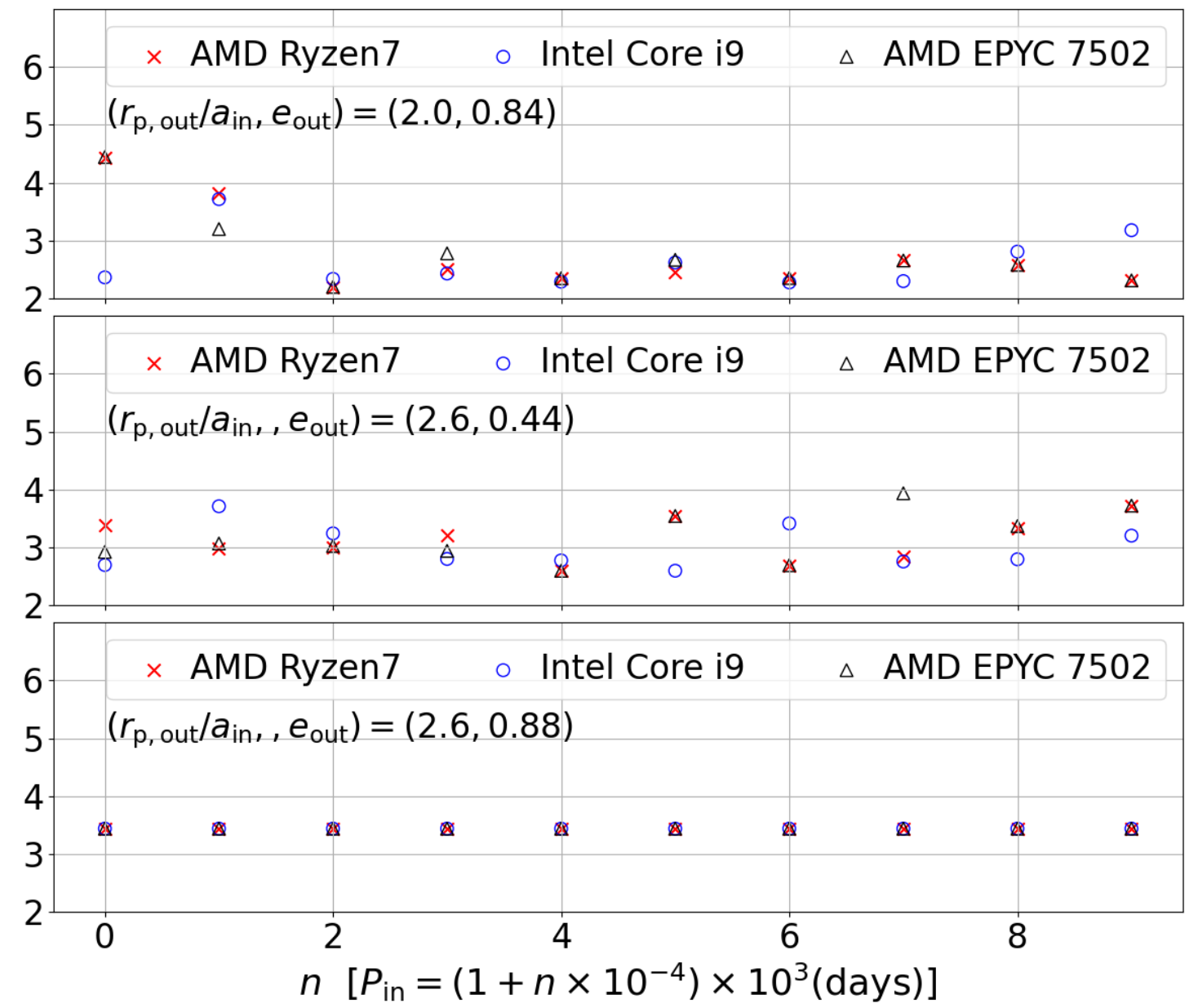}
\end{center}
\caption{Normalized disruption timescales $T_\mathrm{d}/P_\ii$ to the
  tiny change of the input value of the inner orbital period $P_\ii$
  for coplanar prograde triples with $q_{21}=1$ and $q_{23}=5$.
    Top, middle, and bottom panels correspond to $(\rpo/a_\ii,
    e_\oo)=(2.0, 0.84)$, $(2.6, 0.44)$, and $(2.6, 0.88)$.  In the
    left and right panels, the 15-th and 4-th digits of the initial
    values of $P_\ii$ are modified, respectively.
	\label{fig:digit}}
\end{figure*}

Instead, this subsection addresses the question how the disruption
timescale is sensitive to the (tiny) difference of the initial orbital
parameters, given the chaotic nature of the gravitational many-body
systems in general \citep[e.g.,][]{Suto1991}. For this purpose, Figure
\ref{fig:digit} examines quantitatively to what extent the disruption
timescale is affected by the tiny difference of the initial period of
the inner binary.

We select three sets of runs from our fiducial model simulation, and
change the initial value of $P_\ii$ (and therefore that of $\ain$)
alone while keeping the other parameters intact. In practice, we
change the initial input value of $P_\ii=10^3$ days only by its 15-th
digit (left panels) and 4-th digit (right panels), which are the
horizontal axes of Figure \ref{fig:digit}, {\it i.e.,} $n=10^{15}
\Delta P_\ii/P_\ii$ and $n=10^{4} \Delta P_\ii/P_\ii$. We also repeat
the identical run on three different machines; AMD Ryzen7 Core in CfCA
calculation cluster at NAOJ, Intel Core i9 on 16-inch Macbook Pro
2019, and AMD EPYC 7502 in our local computer cluster.  In this
subsection, we stop the runs at $t_\mathrm{int}=4.0\times 10^7P_\ii$
to save computational time.

Let us look at the left panels of Figure \ref{fig:digit} first.  The
bottom panel corresponds to a case that the disruption time is
insensitive to such a small change. The cases plotted in the top and
middle panels, in contrast, indicate that $T_{\rm d}$ varies
significantly among different machines; tiny differences in $P_\ii$
lead to even by a few orders of magnitudes.  This amazing sensitivity
to the initial condition originate from the intrinsically chaotic
nature of the gravitational dynamics of triples. The variability among
different machines is instead due to the limits of floating-point
arithmetic and the differences in optimizing compilers that result in
slightly different machine code being generated. Together with the
extreme sensitivity to initial conditions, these factors cause the
introduction of ``numerical chaos'', when small but different
accumulations of integration errors result in a completely different
evolution of the same system.

We repeated those runs except that we modify only the 4-th digit of
the initial $P_\ii$.  The result is shown in the right panels of
Figure \ref{fig:digit}, which are very similar to the left panels. For
a triple whose $T_{\rm d}$ is insensitive to the 15-th digit of the
input $P_\ii$ (the bottom panels), the value is identical in the three
different machines even if the 4-th digit is modified, implying that
the disruption of the triple seems to proceed in a regular, instead of
chaotic, fashion. We could speculate that the Lyapunov time of such
system is so long that even a relatively large change in the initial
conditions does not result in a different evolution. In contrast, the
range of variance of $T_{\rm d}$ is statistically the same between the
left and right panels for the top and middle examples, suggesting that
their chaotic nature is fairly independent of the fluctuation
amplitude of the input $P_\ii$.  Such a chaotic degree of the unstable
triples is extremely sensitive to the set of orbital parameters in a
complicated fashion, and is difficult to predict. Nevertheless, Figure
\ref{fig:digit} indicates that the disruption timescales from our
current simulations should be understood to vary by one or two orders
of magnitudes, in practice.
  
The disruption timescale $T_\mathrm{d}$ should admit the scaling
relation, equation (\ref{eq:scaling}), with respect to $P_\ii$ and
$(m_1,m_2,m_3)$, as long as non-Newtonian effects are neglected. For
instance, the RW model prediction, equation (\ref{eq:rwmodel}),
respects the scaling.  In reality, however, the numerical results do
not necessarily satisfy the scaling due to the intrinsic chaoticity
described in the above.

\begin{figure*}
\begin{center}
	\includegraphics[clip,width=10cm]{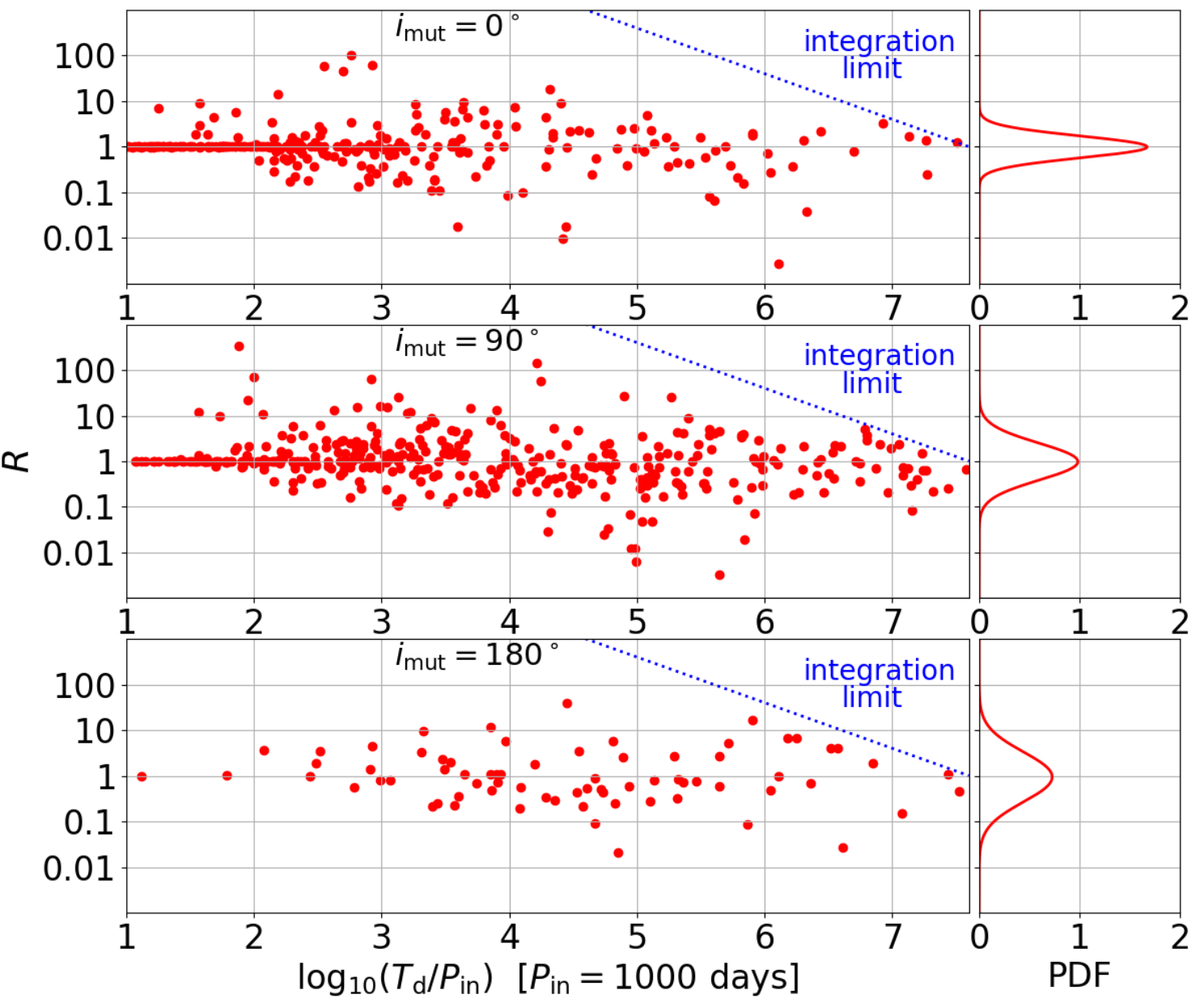}
\end{center}
\caption{Ratio of the normalized disruption timescales for individual
  triples ($q_{21}=1$ and $q_{23}=5$) of $P_\ii= 10^3$ and $10^4$ days
  against those of $P_\ii= 10^3$ days (left), and their corresponding
  probability density function (right).
 \label{fig:chaos_hist}}
 \end{figure*}

In order to see this, we run simulations for a subset of fiducial
triple systems corresponding to Figures \ref{fig:dist_0} to
\ref{fig:dist_180} except that we change the value of $P_\ii$ from
$10^3$ days to $10^4$ days (note that $P_\oo$ is derived from the
other fixed parameters). We define the ratio of the normalized
disruption timescales for individual triples of $P_\ii= 10^3$ and
$10^4$ days:
\begin{eqnarray}
\label{eq:ratio-Td-Pin}
R  = \frac{(T_\mathrm{d}/P_\ii)_{P_\ii=10000 {\rm days}}}
{(T_\mathrm{d}/P_\ii)_{P_\ii=1000 {\rm days}}} .
\end{eqnarray}
Figure \ref{fig:chaos_hist} plots $R$ against
$\log_{10}(T_\mathrm{d}/P_\ii)_{P_\ii=1000 {\rm days}}$ for prograde
(top), orthogonal (middle), and retrograde (bottom) triples with
$q_{21}=1$ and $q_{23}=5$. In order to save the CPU time, we adopt
$t_\mathrm{int}=4\times10^7P_\ii$ in those examples.

The ratios derived from simulations are not exactly equal to unity due
to the chaotic behavior of the triple dynamics, but they distribute
around unity in a statistical sense. For instance, we find that their
distribution is well approximated by log-normal functions as indicated
in the right panels.

In summary, we find that the numerically estimated disruption
timescales are inevitably affected by the chaotic behavior of triple
dynamics.  The timescales may vary one or two orders of magnitudes due
to the differences of the initial phase angles and also to the tiny
changes in the other orbital parameters.  Nevertheless, we conclude
that the disruption timescales are statistically robust, and can be
predicted from the orbital parameters within one or two
order-of-magnitude uncertainty.

\section{Comparison with simulated disruption times and previously
  proposed models: prograde, orthogonal, and retrograde
  orbits \label{sec:comparison}}

Having in mind the scatter of the simulated disruption timescales that
we discussed in subsection \ref{subsec:chaos-scaling}, we present a
detailed comparison of the simulation results against the previous
theoretical models (MA01 and MK20). In particular, we examine the
dependence on the mutual inclination between the inner and outer
orbits.  For definiteness, we focus on the three cases including
coplanar prograde ($i_\mathrm{mut}=0^\circ$), orthogonal
($i_\mathrm{mut}=90^\circ$), and coplanar retrograde
($i_\mathrm{mut}=180^\circ$) triple systems. For each case, we
consider $q_{21}\equiv m_2/m_1 =$ $1$ and $0.1$, and $q_{23}\equiv
m_2/m_3=$$5.0$, $1.0$, and $0.5$. We perform simulations for those 18
orbital configurations in total, by varying $e_\oo$ and $\rpo/a_\ii$
as described in section \ref{subsec:equalmass-sim}.

\subsection{The stability boundary
  in the \texorpdfstring{$e_\oo$ - $\rpo / a_\ii$}{eout - r/a} plane}
\label{sec:staboundary}

\begin{figure*}
\begin{center}
	\includegraphics[clip,width=14.0cm]{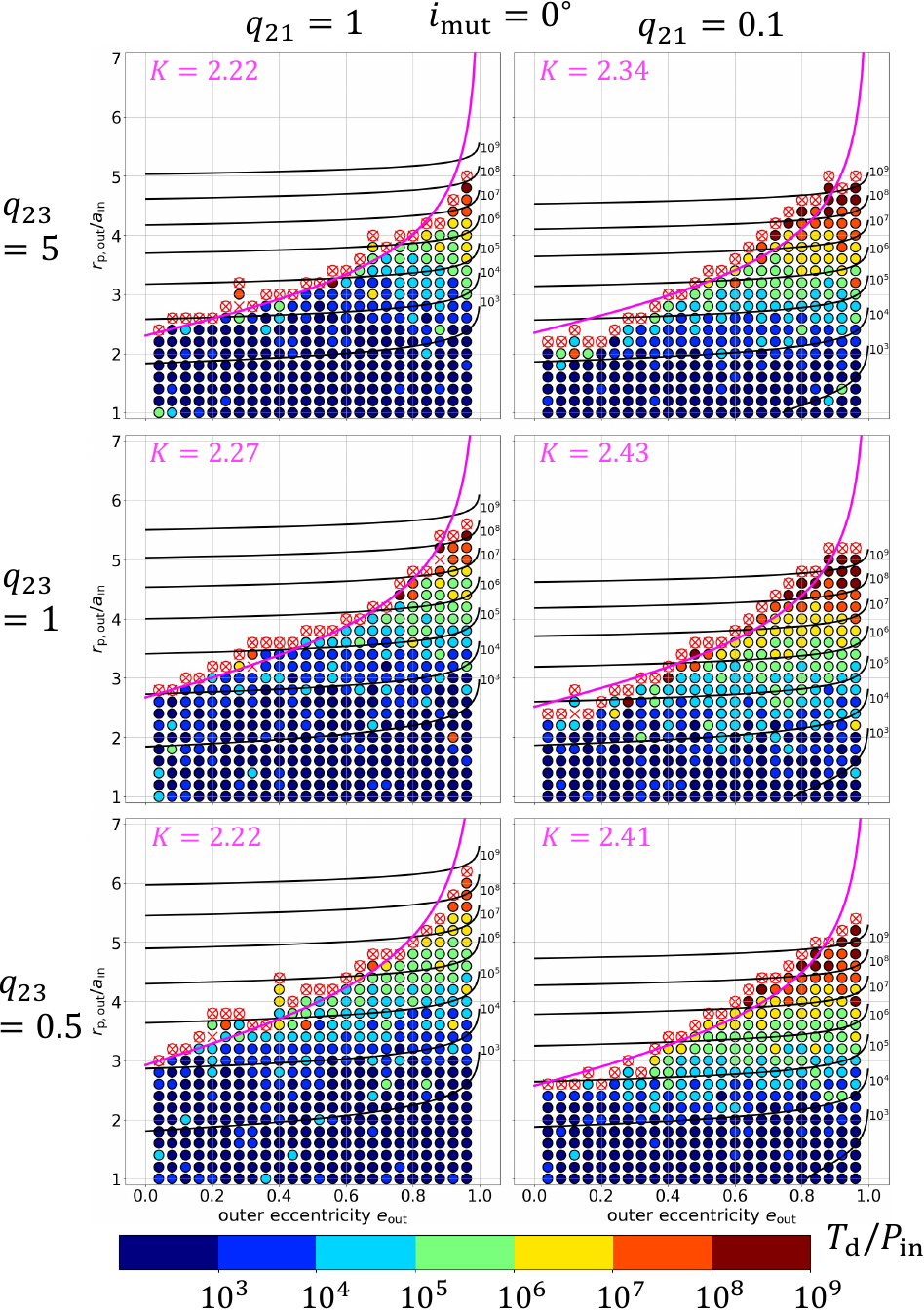}
\end{center}
\caption{Disruption timescales of coplanar prograde triples
  $T_\mathrm{d}/P_\ii$.  Left and right panels correspond to
  $q_{21}=1$ and $0.1$, while top, center, and bottom panels are for
  $q_{23}=0.5$, $1$ and $5$, respectively.
\label{fig:dist_0-q21q23}}
\end{figure*}

\begin{figure*}
\begin{center}
	\includegraphics[clip,width=14.0cm]{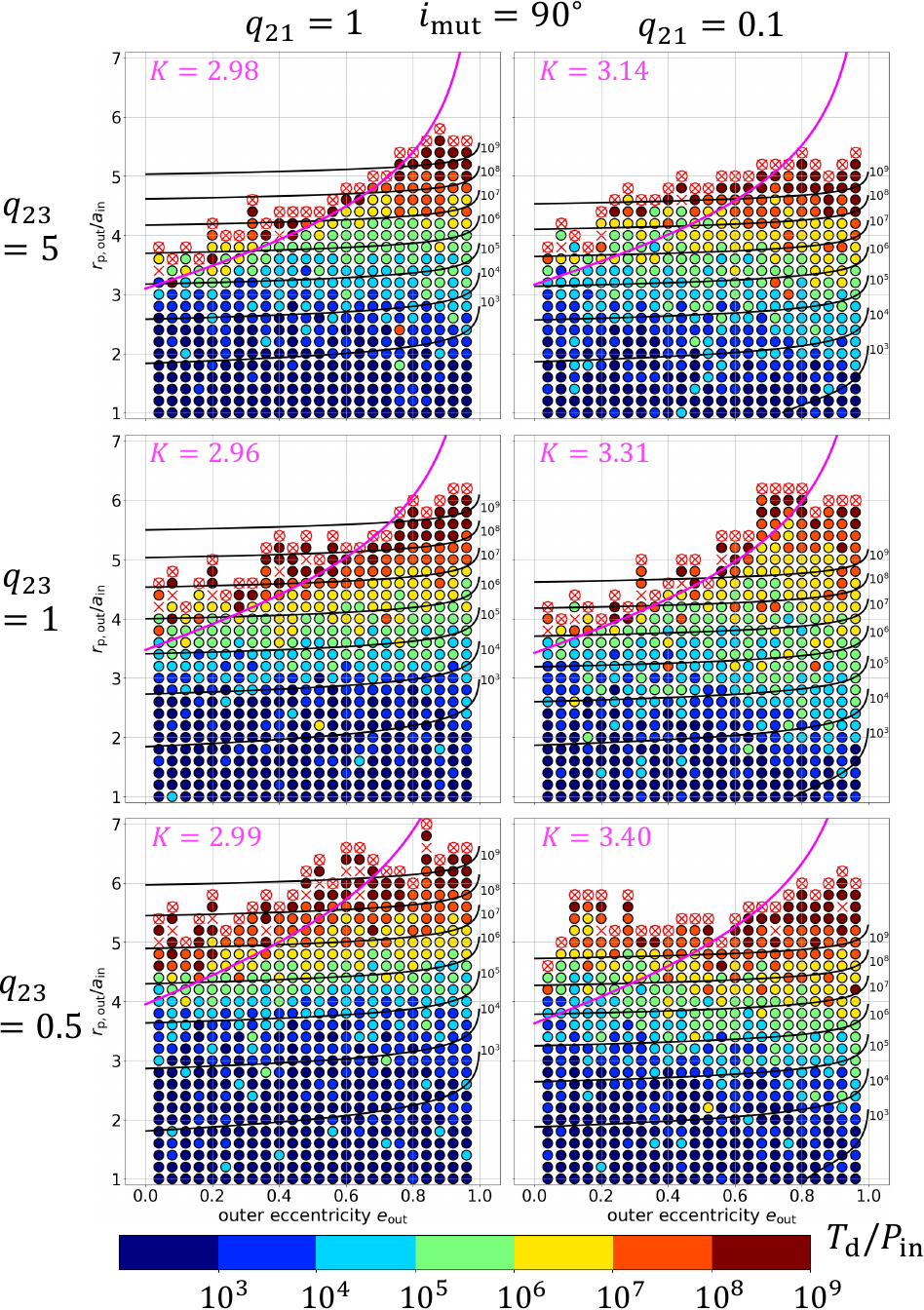}
\end{center}
\caption{Same as Figure \ref{fig:dist_0-q21q23} but for
  initially orthogonal triples.
  \label{fig:dist_90-q21q23}}
\end{figure*}

\begin{figure*}
\begin{center}
	\includegraphics[clip,width=14.0cm]{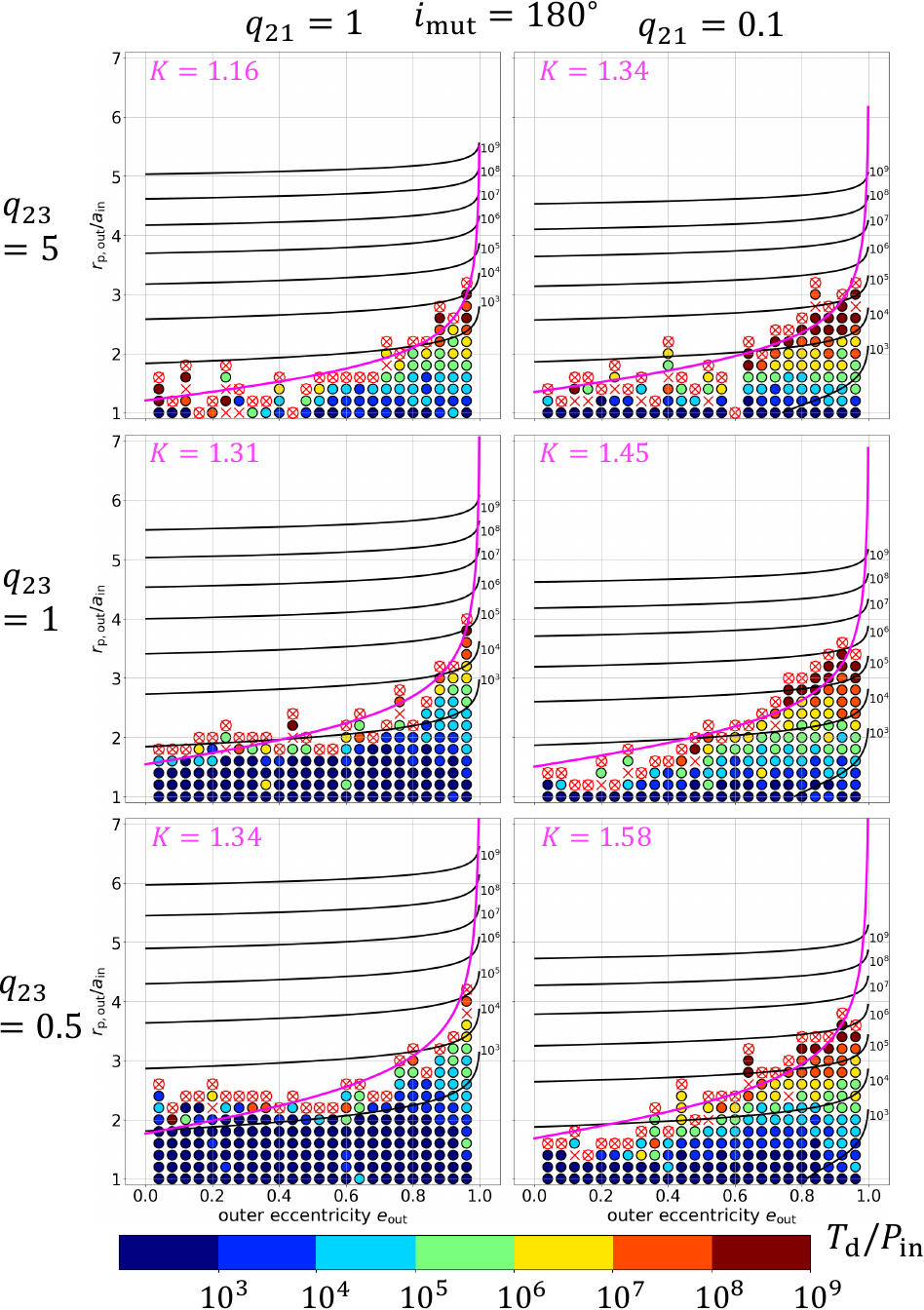}
\end{center}
\caption{Same as Figure \ref{fig:dist_0-q21q23} but for
  coplanar retrograde triples.
  \label{fig:dist_180-q21q23}}
\end{figure*}

Figures \ref{fig:dist_0} to \ref{fig:dist_180} suggest that there is a
fairly sharp boundary of the triple stability on $e_\oo - \rpo/a_\ii$
plane.  The boundary seems to be roughly approximated by shifting
inequality (\ref{eq:MAcriterion}) along the vertical direction. Thus,
we empirically introduce the coefficient $K$ to the MA01 model, and
fit the simulated stability boundary as
\begin{eqnarray}
\label{eq:boundary-model}
\left(\frac{\rpo}{\ain}\right)_{\rm SB} = K(\imut, q_{21}, q_{23})
\left[\left(1+\frac{m_3}{m_{12}}\right)
  \frac{(1+e_\oo)}{\sqrt{1-e_\oo}}\right]^{2/5} .
\end{eqnarray}
The results are shown in Figures \ref{fig:dist_0-q21q23} to
\ref{fig:dist_180-q21q23}, together with the value of $K$.

Although equation (\ref{eq:boundary-model}) is a simple extrapolation
of the MA01 model, its $e_\oo$ dependence reproduces the numerical
results reasonably well by fitting the value of $K$, especially given
the expected scatters discussed in subsection
\ref{subsec:chaos-scaling}. The boundaries between stable and unstable
regions appear to be very sharp especially for $e_\oo<0.3$ and becomes
gentle for $e_\oo>0.8$, at least for $\imut=0^\circ$ and $180^\circ$
(Figures \ref{fig:dist_0-q21q23} and \ref{fig:dist_180-q21q23}).
Unlike inequality (\ref{eq:MAcriterion}), however, the stability
boundary does not monotonically depend on the mutual inclination
$\imut$, and the orthogonal triples are more unstable than the other
two coplanar configurations; see Figure \ref{fig:dist_90-q21q23}.  In
all cases, the disruption timescales for those triples with
${\rpo}/{\ain}<({\rpo}/{\ain})_{\rm SB}$ are mainly determined by the
value of ${\rpo}/{\ain}$, in agreement with the MK20 model.

\begin{figure*}
\begin{center}
	\includegraphics[clip,width=10.0cm]{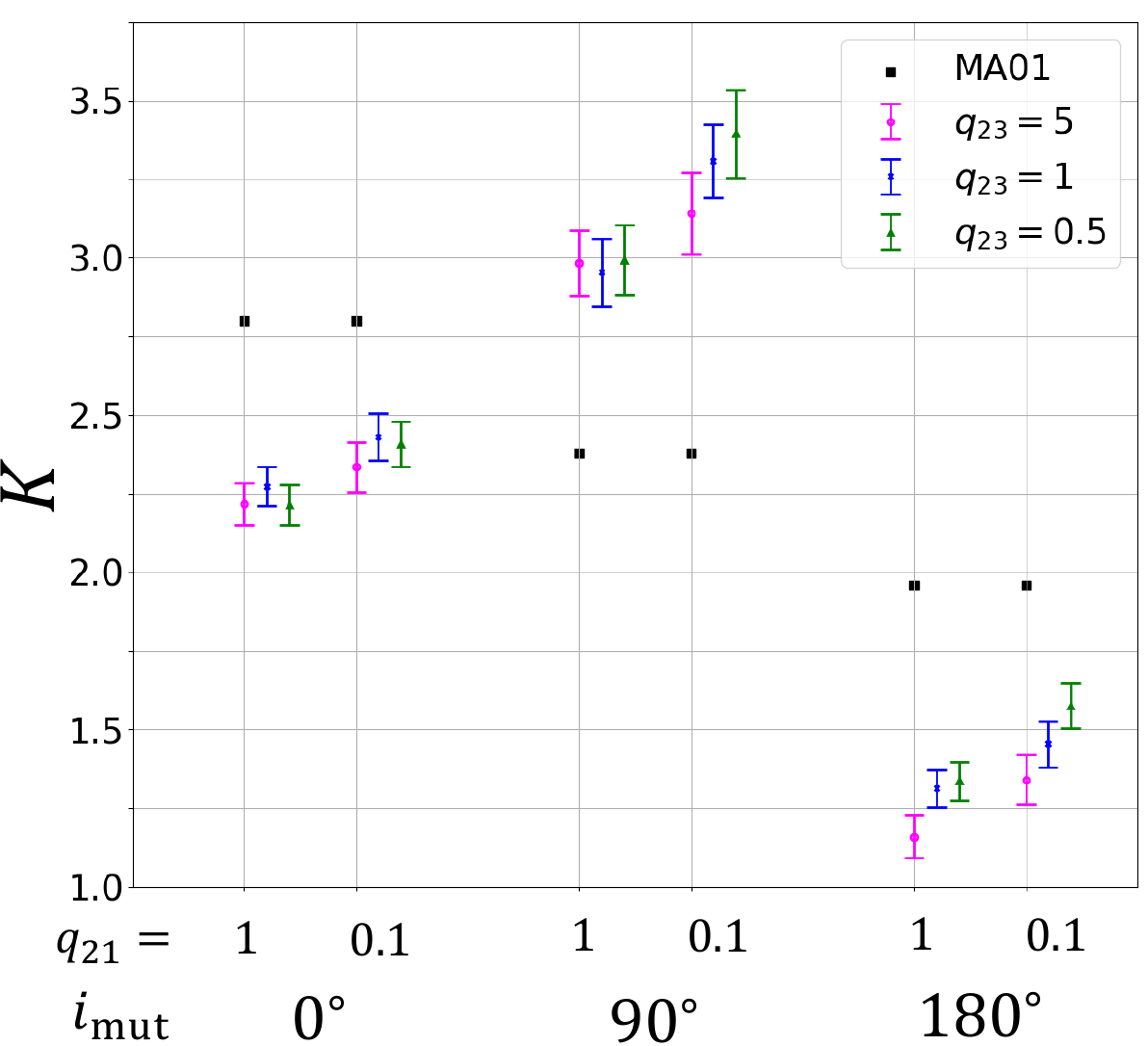}
\end{center}
\caption{The best-fit values of the proportional factor $K(\imut,
  q_{21}, q_{23})$ in equation (\ref{eq:boundary-model}) corresponding
  to Figures \ref{fig:dist_0-q21q23} to
  \ref{fig:dist_180-q21q23}. The quoted error-bars represent
    the $1\sigma$-confidence range. For reference, the values of the
  MA01 model are plotted in black squares.
  \label{fig:Kvalue}}
\end{figure*}

Figure \ref{fig:Kvalue} plots the best-fit values of $K(\imut, q_{21},
q_{23})$, which clearly shows the parameter dependence of the
stability boundaries in Figures \ref{fig:dist_0-q21q23} to
\ref{fig:dist_180-q21q23}. The fit is not so good in some cases,
especially for orthogonal triples. Nevertheless, we perform the
least-square fitting for them, and plot the best-fit values of $K$ in
Figure \ref{fig:Kvalue}, so as to indicate the approximate behavior of
the stability boundary in a qualitative sense. The poor fit for
  orthogonal triples may imply the importance of their complicated
  orbital evolution due to the ZKL oscillation; see section
  \ref{sec:evolution}.  Relative to the MA01 criterion, orthogonal
triples are more unstable and coplanar retrograde triples are more
stable. The dependence on $q_{23}$ is not well represented by the
combination of $(1+m_3/m_{12})$ alone. These differences may likely
come from the fact that their model is based on the simulations up to
$10^2 P_\oo$, which is much shorter than our $t_{\rm int}=10^9
P_\ii$. Of course, it simply reflects the progress of computational
resources over the last two decades, and we would like to emphasize
the amazing insight and quality of the pioneering work by MA01. We
plan to explore the dependence of the dynamical stability boundary on
$\imut$, $q_{21}$ and $q_{23}$ in detail and report the result
elsewhere.

\subsection{Empirical fit to the disruption timescales}
\label{sec:distimescales}

We find that there is a relatively clear boundary between dynamically
stable and unstable regions on $\rpo / a_\ii - e_\oo$ plane. The
boundary is reasonably well approximated by equation
(\ref{eq:boundary-model}), which is a simple generalization of the
MA01 model. In the unstable regions, the disruption timescales seem to
be roughly consistent with the MK20 model. Thus, following equation
(\ref{eq:rwmodel}), we introduce a parameter $x$:
\begin{eqnarray}
\label{eq:x}
x &\equiv& \frac{4\sqrt{2}}{3}
\sqrt{\frac{m_{12}}{m_{123}}}(\log_{10}{e})
     {\left(\frac{r_\mathrm{p,\oo}}{a_\ii}\right)}^{3/2} 
\approx 0.82\sqrt{\frac{1+q_{21}}{1+q_{21}+q_{21}/q_{23}}}
     {\left(\frac{r_\mathrm{p,\oo}}{a_\ii}\right)}^{3/2}.
\end{eqnarray}
Then, equation (\ref{eq:rwmodel}) is rewritten in terms of $x$ as
follows, which clearly implies that the value of $x$ basically
determines the disruption time:
\begin{eqnarray}
\label{eq:rwmodel_x}
\log_{10}{\frac{T_\mathrm{d}}{P_\ii}}
&=& x -\frac{4}{3}\log_{10}x
+ \frac{4}{3}\log_{10}
\left[\left(1+ \frac{1}{q_{21}}+\frac{1}{q_{23}}\right)^{1/2}
\left(1+ q_{21}\right)^2\frac{1}{q_{21}}\right] + \log_{10}(1-e_\oo)^{1/2} \cr
&& \qquad\qquad +\frac{4}{3}\log_{10}\left(\frac{2^{13/4}}{3} \log_{10}e\right).
\end{eqnarray}

\begin{figure*}
\begin{center}
 		\includegraphics[clip,width=12.cm]{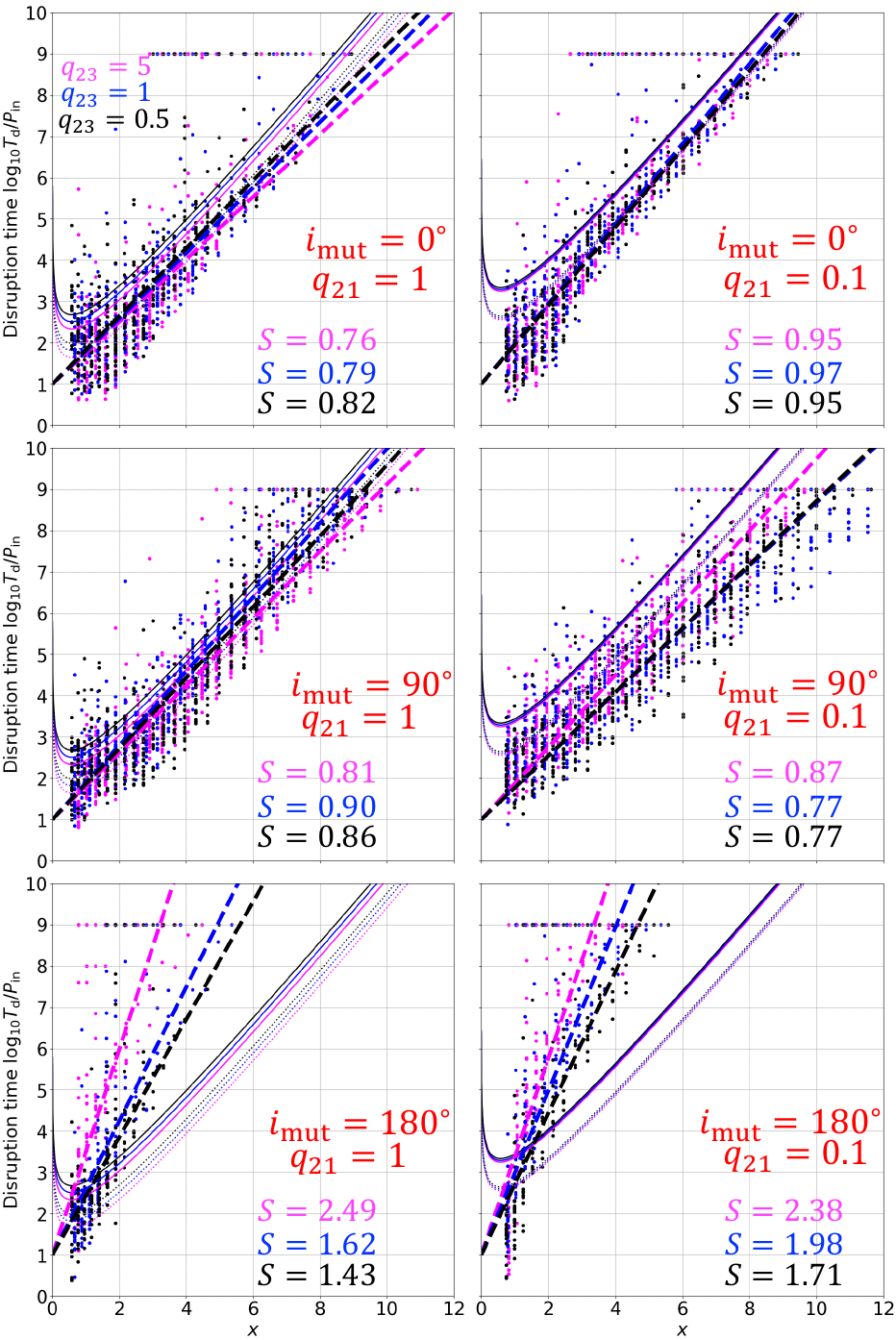}
\end{center}
\caption{Normalized disruption time $T_\mathrm{d}/P_\ii$ distributions
  for the triples in terms of $x$. The magenta, blue, and black dots
  correspond to $q_{23}=$ $5.0$, $1.0$, and $0.5$, respectively. Top,
  middle, and bottom panels show the results for prograde, orthogonal,
  and retrograde cases, respectively. Left and right panels show the
  cases for $q_{21}=1.0$, and $0.1$, respectively. The solid and
  dotted curves denote the corresponding MK20 RW model estimation for
  $e_\oo = 0.04$ and $0.96$, which are the lower and upper limit
  eccentricities in the simulations, respectively.  The dashed lines
  represent the best-fit empirical models. \label{fig:distribution_y}}
 \end{figure*}  

Figure \ref{fig:distribution_y} compares the $T_\mathrm{d}/P_\ii$
distribution from our simulation against the MK20 model; top, middle,
and bottom panels show the prograde ($i_\mathrm{mut}=0^\circ$),
orthogonal ($i_\mathrm{mut}=90^\circ$), and retrograde cases
($i_\mathrm{mut}=180^\circ$), while left and right panels are for
equal-mass ($q_{21} = 1.0$) and unequal-mass inner binaries.  The
results with different $q_{23}$ are plotted in different colors:
$q_{23}=5.0$ (magenta), $1.0$ (blue), and $0.5$ (black). Thin solid
and dotted lines indicate the MK20 model predictions, equation
(\ref{eq:rwmodel_x}), for $e_\oo=0.04$ and $0.96$, respectively, so as
to bracket the dependence on $e_\oo$.  Filled circles in different
colors correspond to our simulation data in Figures
\ref{fig:dist_0-q21q23} to \ref{fig:dist_180-q21q23}. The sequence of
data located at $T_\mathrm{d}/P_\ii=10^9$ corresponds to triples that
are dynamically stable up to our integration time $t_\mathrm{int}=10^9
P_\ii$, and thus are not considered in the discussion below.

Let us examine more quantitatively Figure \ref{fig:distribution_y}.
Coplanar prograde triples (top panels) do not exhibit noticeable
dependence on $q_{23}$ in the range $0.5$ - $5.0$, which is consistent
with the very weak dependence on $q_{23}$ that the MK20 model
predicts. Furthermore, unequal-mass inner binaries ($q_{21}=0.1$) tend
to be slightly more stable than equal-mass ones, which is in a
qualitative agreement as well with the MK model.

Orthogonal triples (middle panels) exhibit no strong dependence on
$q_{23}$ either, given the large scatters of simulated data, while
unequal-mass inner binaries seem to be more stable for less massive
tertiaries (with $q_{23}=5$) in contrast to the MK20 model.  This may
indicate the importance of other processes than the RW energy
transfer, such as the ZKL oscillation up to octupole interaction. In
section \ref{sec:evolution}, we show one typical orbital evolution
under the ZKL oscillation.

Coplanar retrograde triples (bottom panels), on the other hand, are
significantly more stable than the above two cases, and behave very
differently relative to the MK20 model, as already indicated in
Figures \ref{fig:dist_0-q21q23} to \ref{fig:dist_180-q21q23}.  The
fact that simply reversing the motion of the outer body changes
significantly the disruption time distribution is not so intuitive,
and we will discuss further in the next section.

Since we have not yet identified physical processes that explain the
orbital configuration dependence on the disruption time, we introduce
a simple empirical model, and determine the best-fit parameters to
summarizes our current results.  For that purpose, we consider the
following parameterization:
\begin{eqnarray}
\label{eq:fit-sxC}
  \log_{10}{\left(\frac{T_\mathrm{d}}{P_\ii}\right)}_\mathrm{fit}
  = Sx + C,
\end{eqnarray}
where the dominant term in the MK20 RW model corresponds to $S=1$.
 
\begin{figure*}
\begin{center}
\includegraphics[clip,width=10.0cm]{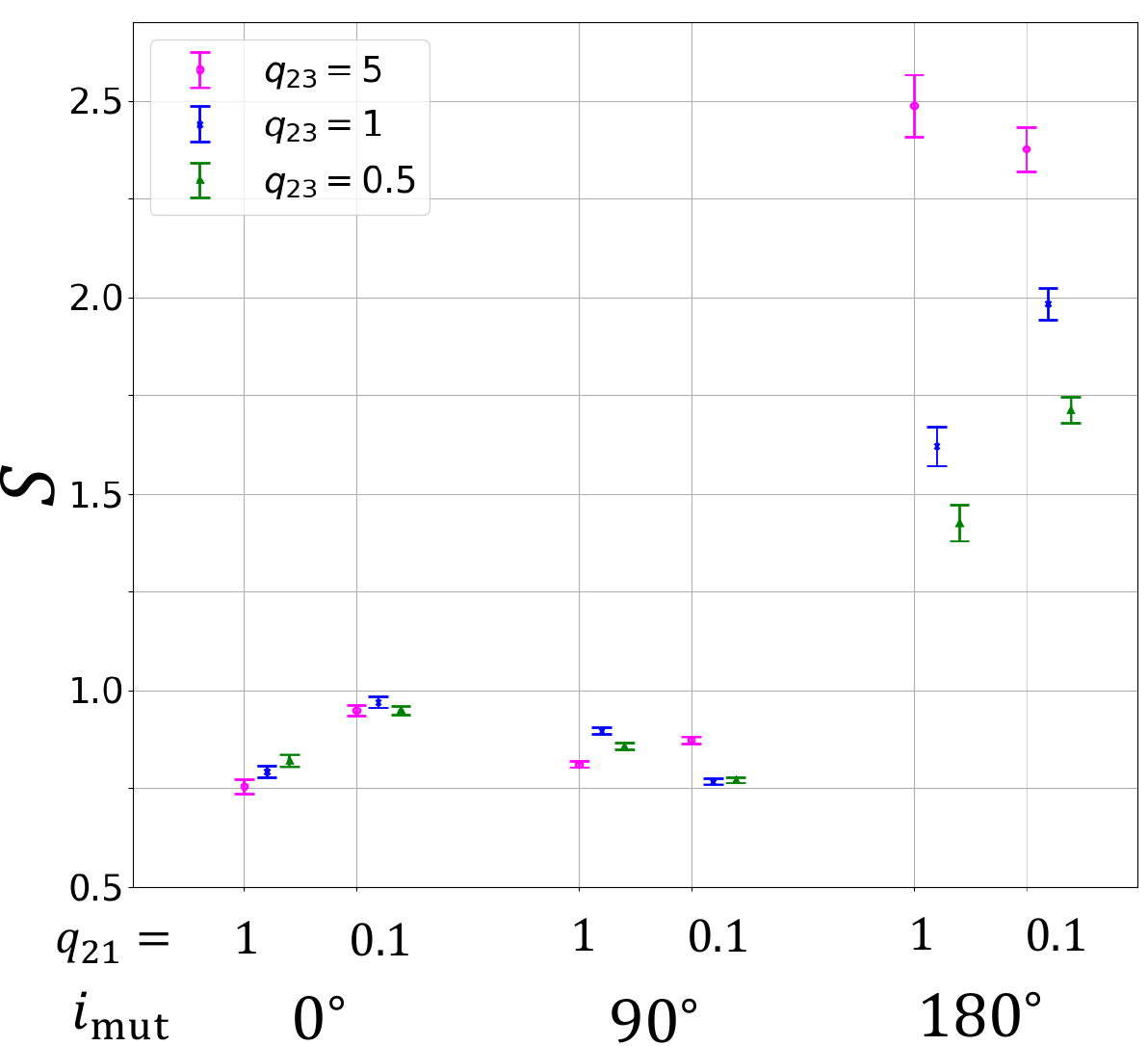}
\end{center}
\caption{The best-fit values for $S$ with their
    $1\sigma$-confidence range.
\label{fig:fitting}}
\end{figure*}

We first attempted to fit the results of Figure
\ref{fig:distribution_y} by varying $S$ and $C$, and found that most
of the results are fitted with $C$ close to unity. Given the scatters
of the disruption timescales due to the initial phase angles and
chaotic behavior of the triples, therefore, we decided to adopt $C=1$
and fit with $S$ alone.  Thick-dashed lines in Figure
\ref{fig:distribution_y} represent the best-fit models corresponding
to equation (\ref{eq:fit-sxC}) with $C=1$.  Figure \ref{fig:fitting}
shows the best-fit value and the associated 1$\sigma$-confidence for
$S$.  While equation (\ref{eq:fit-sxC}) is a very rough approximation,
it depicts our simulation results at least qualitatively, and is
helpful to understand their basic behavior in a simple way.

\section{Orbital evolution of hierarchical triples
  with different mutual inclinations \label{sec:evolution}}

We have shown that the disruption timescales are very sensitive to the
mutual inclination between the inner and outer orbits.  In order to
understand their disruption processes, we first show how various
orbital elements in our numerical simulations evolve differently
according to the mutual inclination.  Then, we explain why the
orthogonal triples are more unstable, and the retrograde
  triples are more stable than the coplanar triples, applying the
energy transfer model between the two orbits for a parabolic encounter
approximation by \citet{Roy2003}.

\subsection{Evolution of orbital elements, energies,
   and angular momenta \label{subsec:orbit-evol}}

\begin{figure*}
\begin{center}
	\includegraphics[clip,width=14cm]{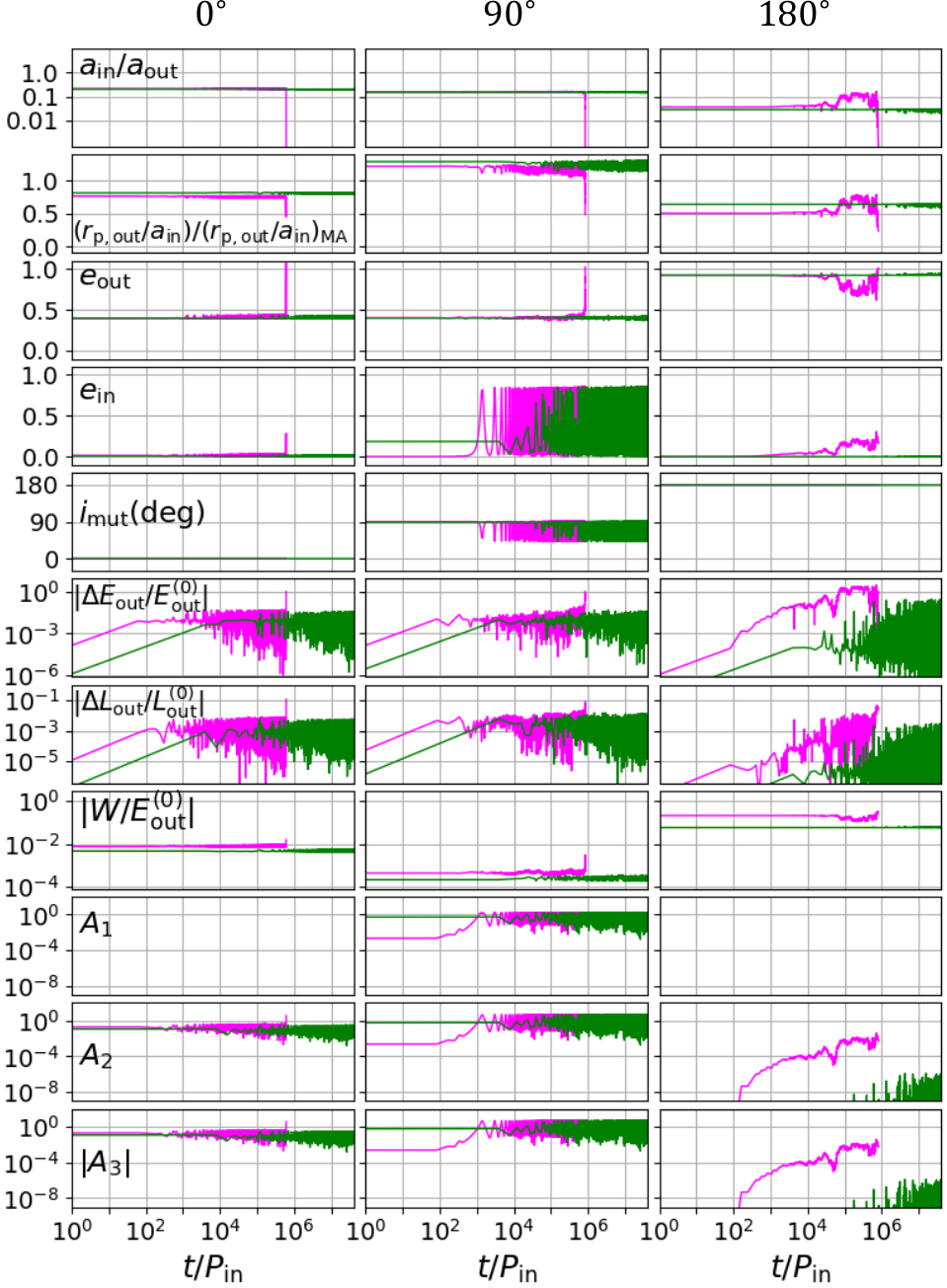}
\end{center}
\caption{Examples of evolution of hierarchical triples for
    $q_{21}=1$ and $q_{23}=5$. The initial orbital parameters are
    $(\rpo/a_\ii, e_\oo, \imut)=(2.8,0.4, 0^\circ)$ (magenta in the
    left panels), $(3.0,0.4, 0^\circ)$ (green in the left panels),
    $(3.8,0.4,90^\circ)$ (magenta in the middle panels),
    $(4.0,0.4,90^\circ)$ (green in the middle panels),
    $(2.2,0.92,180^\circ)$ (magenta in the right panels), and
    $(2.8,0.92,180^\circ)$ (green in the right panels).
\label{fig:orbital-evolution}}
\end{figure*}

Evolution of orbital elements, energies and angular momenta for
  hierarchical triples with $q_{21}=1$ and $q_{23}=5$ is plotted in
  Figure \ref{fig:orbital-evolution}. Left, middle and right panels
  are for the coplanar prograde, orthogonal, and coplanar retrograde
  triples. In each panel, magenta and green curves represent the
  evolution of systems that disrupt around $10^6 P_\ii$
  and remain stable at $4\times 10^7 P_\ii$, respectively.

From top to bottom panels, we plot the semi-major axis ratio of the
inner and outer orbits $a_\ii/a_\oo$, the ratio of
$(r_\mathrm{p,out}/a_\ii)$ and the MA01 threshold
$(r_\mathrm{p,out}/a_\ii)_\mathrm{MA}$, the outer eccentricity
$e_\oo$, the inner eccentricity $e_\ii$, the mutual inclination
$i_\mathrm{mut}$, the fractional variation of the total energy and
angular momentum of the outer orbit $|\Delta E_\oo/E^{(0)}_\oo|$ and
$|\Delta L_\oo/L^{(0)}_\oo|$, $|W/E_\oo|$, and finally $A_i$ ($i=1,2,$
equations~\ref{eq:A1}--\ref{eq:A2}) and $|A_3|$
(equation~\ref{eq:A3}).

Both coplanar triples remain coplanar for the entire evolution,
  and their inner eccentricities are relatively small. In contrast, the
  initially orthogonal triple changes the mutual inclination between
  $40^\circ$ and $90^\circ$ due to the ZKL effect, which is
  accompanied by the large periodic oscillation of $e_\ii$. As we will
  show in the next subsection, this is why the initially orthogonal
  triples are more unstable than the coplanar cases.

After a certain period of the energy transfer between the inner
  and outer orbits, $a_\ii/a_\oo$ drops and $e_\oo$ is enhanced
  suddenly for unstable cases, leading to an ejection of
  the tertiary.

\subsection{Dependence of the energy transfer on the mutual inclination}
\label{sec:mutincenergytrans}

The fact that the disruption timescales are sensitive to the mutual
inclination angle between the inner and outer orbits may be understood
from the model by \citet{Roy2003}, which is briefly summarized in
section \ref{subsec:roy2003}. Specifically, equations
(\ref{eq:dE-W-F}) and (\ref{eq:F-A}) indicate that the energy exchange
between the inner orbit and a parabolic perturber is approximately
estimated as
\begin{eqnarray}
  \Delta E_\oo = FW = (\sqrt{2}A_1\sin{\phi}+
  2A_2\sin{\phi}\cos{2\Omega} + 2A_3\cos{\phi}\sin{2\Omega}) W,
\end{eqnarray}
where $W$ is given by equation (\ref{eq:def-W}).

Let us assume that the energy transfer between the inner and
  outer orbits in our hierarchical triples is approximated by such an
  encounter, which happens periodically at a pericenter of the
  eccentric outer orbit.

\begin{figure*}
\begin{center}
\includegraphics[clip,width=7.7cm]{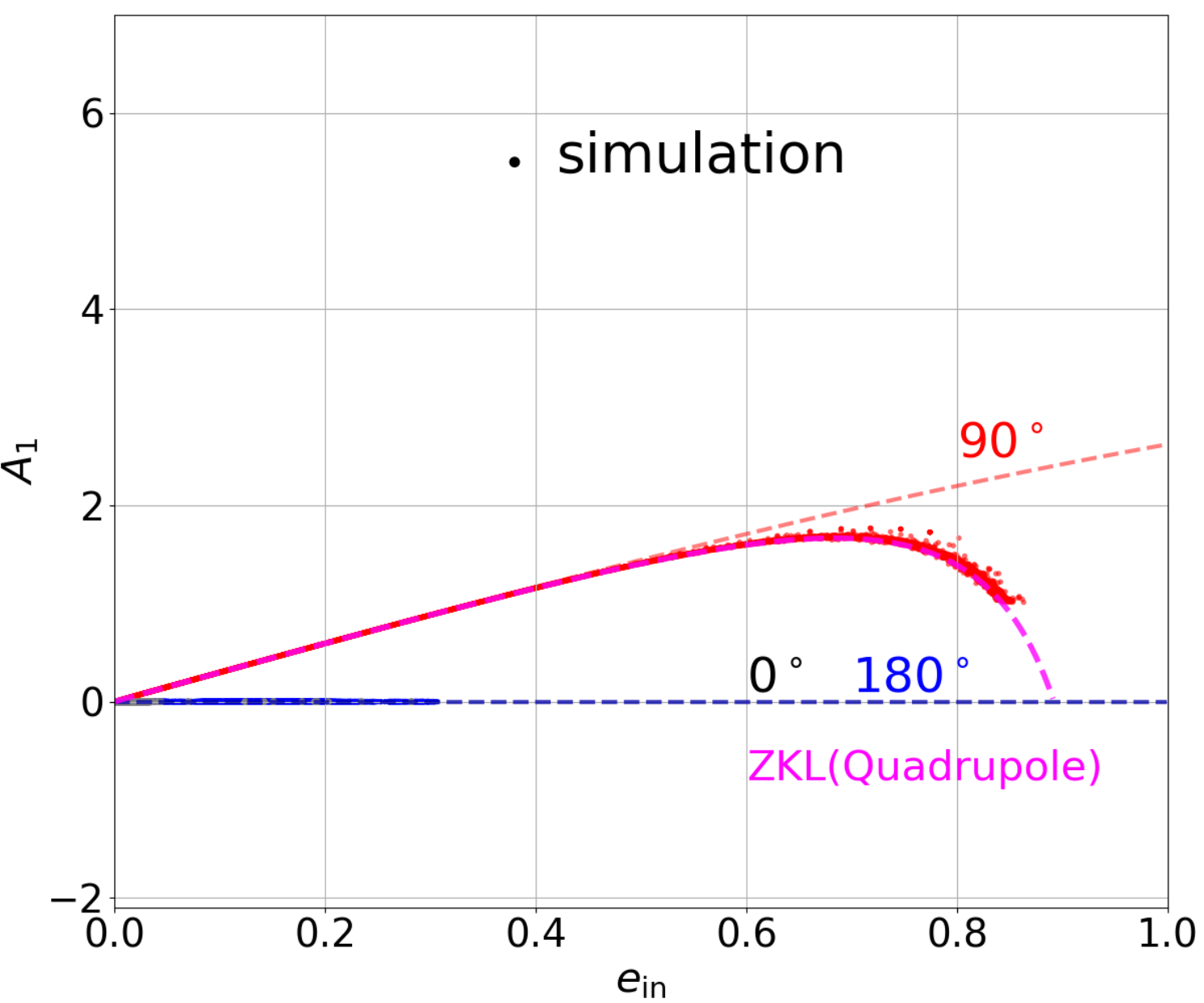}
\hspace{100pt}
\includegraphics[clip,width=7.7cm]{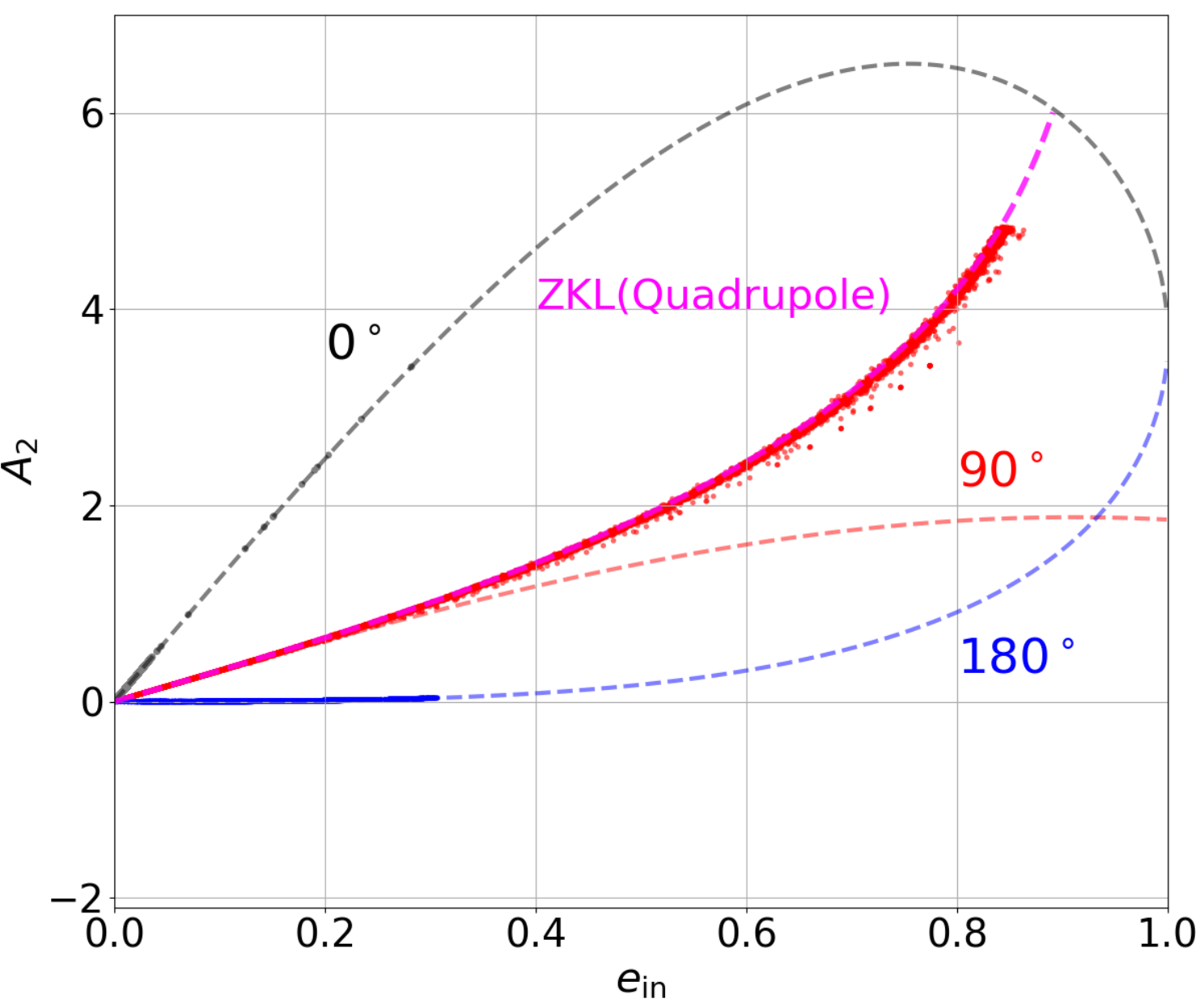}
\hspace{100pt}
\includegraphics[clip,width=7.7cm]{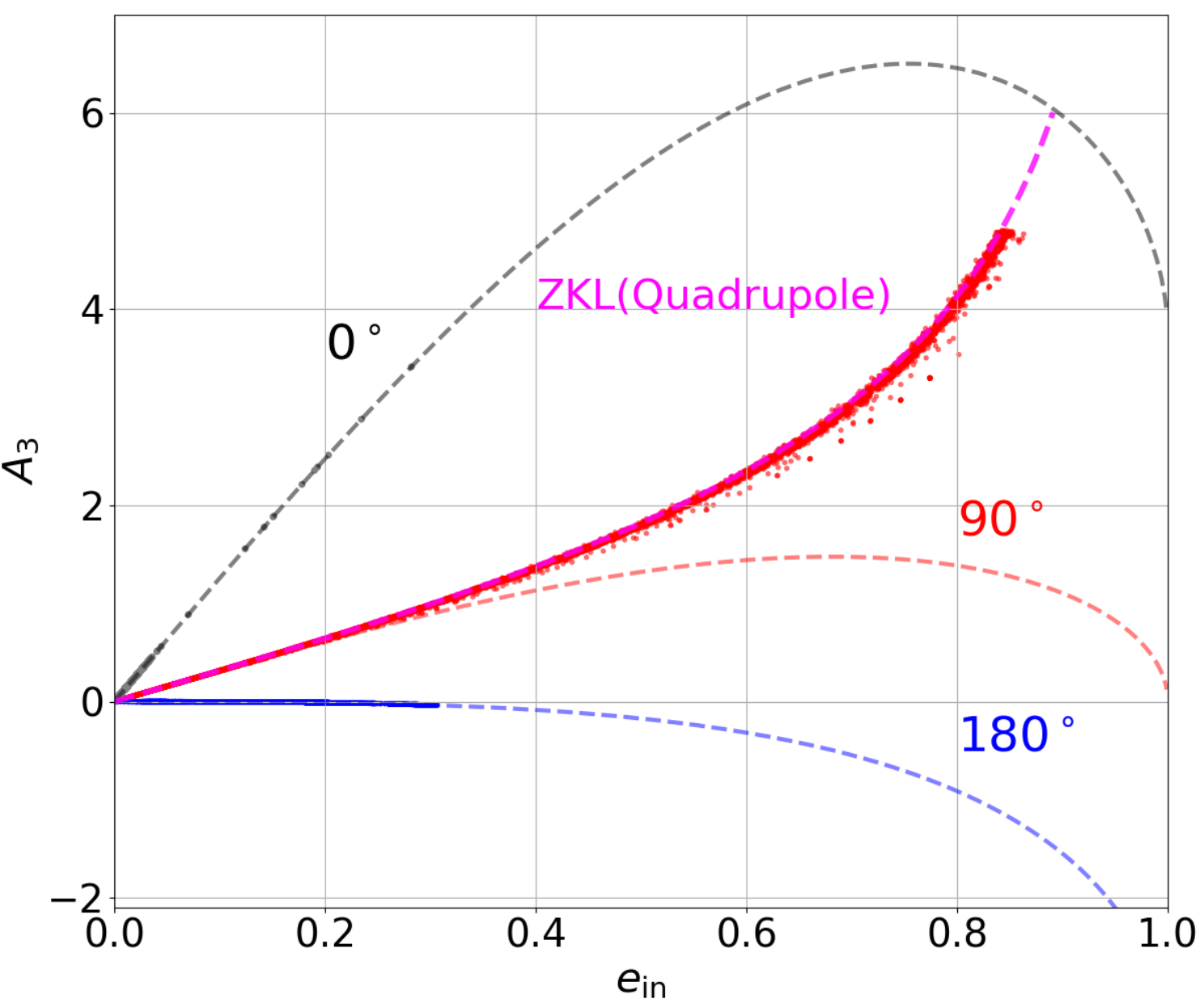}
\end{center}
\caption{Energy transfer coefficients $A_1$ (top), $A_2$
    (middle), and $A_3$ (bottom) against $e_\ii$ for a parabolic
    encounter with $\imut=0^\circ$ (black), $90^\circ$ (red), and
    $180^\circ$ (blue). Dots plotted with the same color represent
    $\sim 10^4$ snapshots from the simulated trajectories
    corresponding to those shown in Figure \ref{fig:orbital-evolution}
    for $(\rpo/a_\ii, e_\oo, \imut)=(2.8,0.4, 0^\circ)$,
    $(3.8,0.4,90^\circ)$, and $(2.2,0.92,180^\circ)$. The curves
    labelled ZKL show the trajectory along $e_\ii=e_\ii(\imut)$ from
    equation (\ref{eq:imut-ein-ZKL}). Note that the sequences of dots
    in each panel may look like solid curves, indicating they
    reproduce well the expected relation between $\imut$ and
    $e_\ii$.} \label{fig:A1A2A3}
\end{figure*}

The coefficients $A_1$ to $A_3$ are explicitly given by
  equations (\ref{eq:def-A1}) to (\ref{eq:def-f4}), and plotted in
  Figure \ref{fig:A1A2A3} against $e_\ii$ for $\imut=0^\circ$ (solid),
  $90^\circ$ (dotted), and $180^\circ$ (dashed).  Expanding those
  equations with respect to $e_\ii$, one obtains their leading-order
  expressions as
\begin{eqnarray}
    \label{eq:A-approx}
  A_1 \approx (2^{3/4}\sqrt{\pi} \sin^2 \imut) e_\ii
  \qquad
  A_2 = A_3 \approx \frac{3\sqrt{\pi}}{2^{3/4}}(1+\cos \imut)^2 e_\ii .
\end{eqnarray}

Equation (\ref{eq:A-approx}) and three curves in Figure
\ref{fig:A1A2A3} seem to suggest that orthogonal triples are more
stable than coplanar prograde triples if compared at the same value of
$e_\ii$.  In reality, however, our simulation runs indicate that the
orthogonal triples are the most unstable.  This is due to the fact
that the inner eccentricity significantly varies for orthogonal
triples due to the ZKL oscillation, thus we have to take account of
the evolution of $e_\ii$ as well. From the conservation of the total
angular momentum in the ZKL effect, one can derive the relation
\citep[e.g.][]{Naoz2016}:
\begin{eqnarray}
\label{eq:imut-ein-ZKL}
\cos\imut = \frac{m_1 m_2 m_{123}^{1/2}}{2 m_{12}^{3/2} m_3}
\left(\frac{a_\ii}{a_\oo}\right)^{1/2}
\frac{e_\ii^2}{(1-e_\ii^2)^{1/2}(1-e_\oo^2)^{1/2}}
\end{eqnarray}
for triples initially with $e_\ii\approx 0$ and $\imut=90^\circ$. In
the quadrupole approximation (or in the case of equal-mass inner
binaries that do not have the octupole term), $e_\oo$ and
$a_\ii/a_\oo$ are constant, and equation (\ref{eq:imut-ein-ZKL})
directly predicts $\imut$ as a function of $e_\ii$. Dots in Figure
\ref{fig:A1A2A3} are sampled from the orbital evolution of the models
shown in Figure \ref{fig:orbital-evolution}, and are in good agreement
with the prediction based on equation (\ref{eq:imut-ein-ZKL});
orthogonal triples exhibit a wide range of $e_\ii$, while in the other
two coplanar cases, $\imut$ is constant and $e_\ii$ stays smaller than
0.3 at most. Thus, the behavior of $A_i$ along the real orbital
evolution of the three models nicely explains why the orthogonal
triples are the most unstable.

\section{Summary and conclusion \label{sec:summary}}

This paper has examined the stability of hierarchical triple systems
using direct $N$-body simulations without adopting a secular
perturbation assumption. We have estimated their disruption timescales
in addition to the mere stable/unstable criterion, with particular
attention to the mutual inclination between the inner and outer
orbits. The simulation results have been intensively compared with the
previous models including the dynamical stability criterion by
\citet{Mardling1999,Mardling2001}, and the random walk disruption
model by \citet{Mushkin2020} among others.  Our main findings are
summarized as follows.
\begin{enumerate}
\item We confirmed that the dynamical stability criterion pioneered by
  \citet{Mardling1999,Mardling2001} captures the basic trend of our
  simulation results.  Since we performed our simulation runs
  significantly longer than their previous ones, however, we are able
  to improve the accuracy of the overall amplitude of the stability
  boundary. Especially, we found that the stability boundary is very
  sensitive to the mutual inclination in a more complicated fashion
  than the empirical fit by \citet{Mardling2001}; coplanar retrograde
  triples and orthogonal triples are much more stable and unstable,
  respectively than coplanar prograde triples
  (section~\ref{sec:staboundary}).
\item We computed the disruption timescales of triples satisfying the
  stability condition up to $10^9 P_\ii$. The timescales follow the
  scaling predicted by \citet{Mushkin2020}, especially with high
  $e_\oo$ where the random walk assumption is supposed to be valid.
  We obtained an improved empirical fit to the disruption timescales,
  which also indicated the coplanar retrograde triples are
  significantly more stable than the previous prediction
  (section~\ref{sec:distimescales}). We successfully explained this
  tendency and also why the initially orthogonal triples are the most
  unstable, applying the energy transfer model for a parabolic
  encounter developed by
  \citet[][section~\ref{sec:mutincenergytrans}]{Roy2003}.
\item 
  We explicitly demonstrated that the disruption timescales, and even
  the disruption occurrence, are dependent on the initial values of
  phase angles including the mean anomalies and apsidal orientations
  of the inner and outer orbits, and vary within a few orders of
  magnitudes (section~\ref{subsec:chaos-scaling} and
  appendix~\ref{sec:Td-phase}). Additionally, variations of these
  angles can even turn a stable system into an unstable one and vice
  versa. This is somewhat expected, but can be evaluated with the
  long-term direct $N$-body runs without resorting to the secular
  perturbation approximation, thanks to the recently developed
  $N$-body integrator {\tt TSUNAMI} \citep[][Trani et al., in
    prep.]{Trani2019,trani2022b}. Perhaps more surprising is the high
  sensitivity to the tiny change of the initial parameters such as
  $P_\ii$; even the level of the numerical truncation error seems to
  change the disruption timescales by a couple of orders of
  magnitude. This reflects the intrinsic chaotic nature of the
  dynamics of the hierarchical triple systems, implying that the
  disruption timescales estimated by the present paper needs to be
  understood in a statistical sense.
\end{enumerate}

Hierarchical triples have been proposed as one possible channel to
accelerate the merger of compact objects via the ZKL oscillation
\citep[e.g.][]{Liu2018,Trani2021}. Our result shows that the
orthogonal triples, for which the ZKL oscillation are highly
efficient, are generally more unstable than previously thought
\citep{Mardling2001}. Therefore, the validity of this channel needs to
be considered with the constraint on the disruption timescales as
well.

Retrograde triples may be also important as feasible targets of survey
for triple systems that include compact inner binaries, as discussed
in our previous studies \citep[e.g.][]{Hayashi2020}. Such triples may
form via few-body encounters in star clusters, and the predicted
mutual inclinations have uniform distribution in
$\cos{i_\mathrm{mut}}$ \citep[e.g.][]{Fragione2020,Trani2021}. Thus,
once retrograde triples form via this process, they are expected to
stay in a stable configuration for a longer time than the coplanar
prograde triples.

In this paper, we only take account Newtonian gravity and neglect GR
corrections. As discussed in the appendix~\ref{sec:grprecession}, GR
corrections may play an important role only for specific orbital
configurations. However, GR correction, especially GW emissions, are
very important to study different processes leading the disruption of
triples, such as inner binary mergers. We plan to consider this
process separately, which will be reported elsewhere.

Finally, we would like to note that hierarchical triples include a
variety of systems. For example, stars with a binary planet and triple
black holes are also interesting systems, although (and exactly
because) they have not been discovered yet. Extrasolar binary planets
may be detected eventually via transit observations
\citep[e.g.][]{Lewis2015}. In order to model the orbital stability of
such systems, and study the feasibility of detection, it is important
to consider the disruption timescales of such systems, and extend
stability models for triple systems to a variety of orbital
configuration, including mutual inclination, mass ratios, and other
orbital parameters.

After completing the draft of the present paper, we noticed a very
recent preprint by \citet{Vynatheya2022}. They improved the stability
criterion by \citet{Mardling2001}, applied machine learning to
distinguish between the stable and unstable hierarchical triples,
analogously to \citet{Lalande2022}, but did not examine their
disruption timescales as we extensively discussed in the present
work. The quantitative comparison with their result will be presented
elsewhere.

\section*{Acknowledgments}

T.H. and Y.S. are grateful to Hideki Asada for his hospitality during
their visit at Hirosaki University, where part of the present work was
performed.  Y.S. also thanks Taksu Cheon for his hospitality at Kochi
University of Technology. The numerical simulations were carried out
on the local computer cluster \texttt{awamori}, and the general
calculation server from Center for Computational Astrophysics (CfCA),
National Astronomical Observatory of Japan (NAOJ).  This work is
supported partly by the Japan Society for the Promotion of Science
(JSPS) Core-to-Core Program “International Network of Planetary
Sciences”, and also by JSPS KAKENHI grant Nos. JP18H01247 and
JP19H01947 (Y.S.), JP21J11378 (T.H.), and JP21K13914
(A.A.T). T.H. acknowledges the support from JSPS fellowship.

\appendix

\section{Disruption timescales for triples: dependence
    on initial phase angles}
\label{sec:Td-phase}

\begin{figure*}
 	\begin{center}
	\includegraphics[clip,width=8cm]{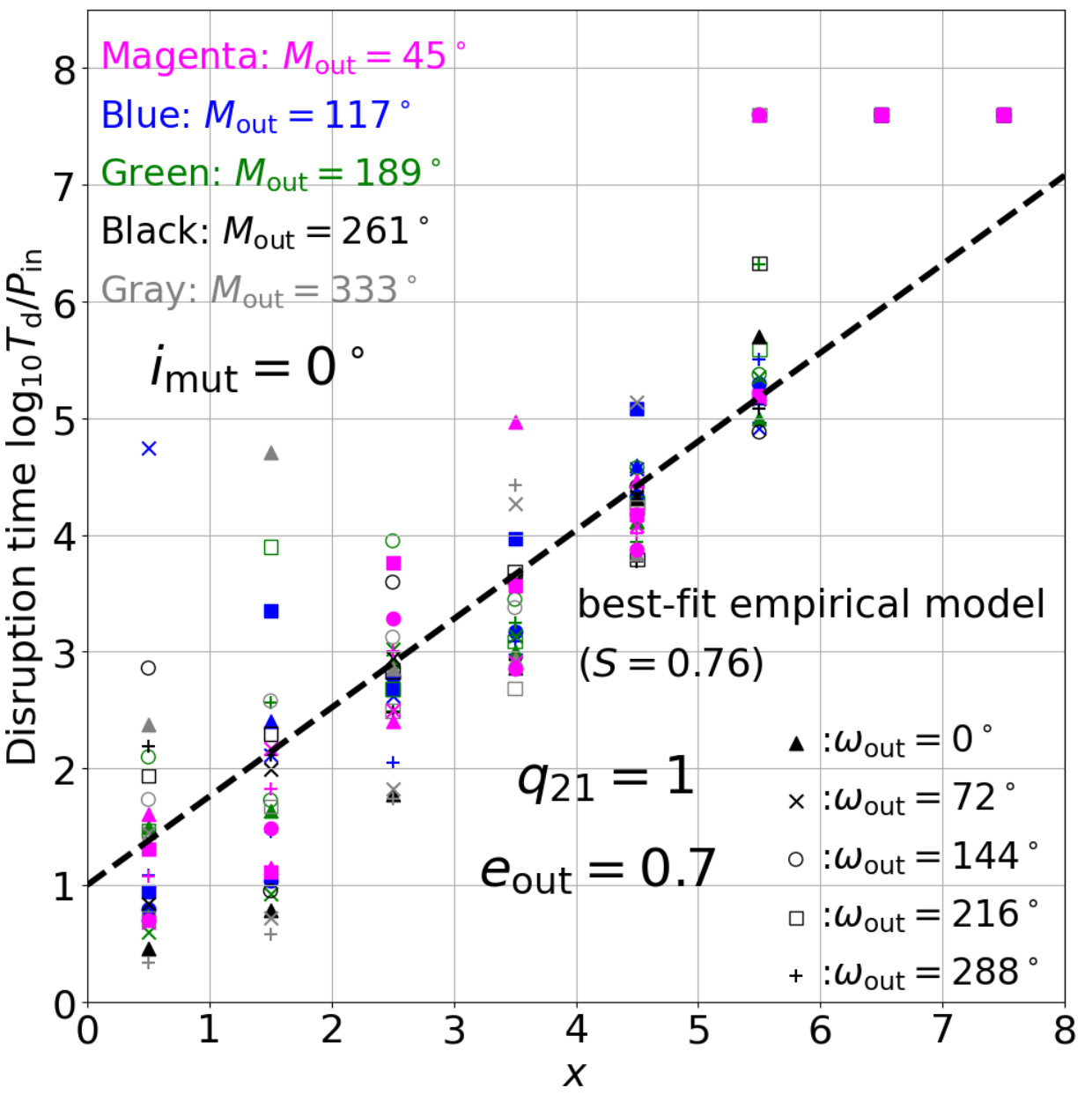}
 	\end{center}
\caption{Normalized disruption timescale ($T_\mathrm{d}/P_\ii$)
  distribution for the coplanar prograde and equal-mass triple against
  $x$ for $q_{21}=1$, $q_{23}=5$, and $e_\oo=0.7$.  For reference, the
  best-fit empirical model from Figure \ref{fig:distribution_y} is
  plotted in the dashed line.}
 	\label{fig:phase_check}
 \end{figure*}

We computed the disruption timescales of triples by fixing the initial
phase angles, but we expect that varying them would add further
variations and scatters in the resulting distribution. In order to
examine the effect of initial phases, we repeated the runs for
coplanar prograde triples with $q_{21}=1$, $q_{23}=5$, and $e_\oo=0.7$
but varying the mean anomaly and pericenter argument of the tertiary,
$M_\oo$ and $\omega_\oo$. The result is shown in Figure
\ref{fig:phase_check}.  For each value of $x= 0.5 + i$ $(i=0$--$7)$,
we compute $25$ realizations by varying the initial outer mean anomaly
$M_\oo$ and the pericentre argument $\omega_\oo$: $(\omega_\oo,
M_\oo)=(72m,45+72n)$ deg, where $m$ and $n$ runs over $0$--$4$.  For
simplicity, we fix the inner angles $M_\ii=30^\circ$ and
$\omega_\ii=180^\circ$. In order to save the CPU time, we adopt
$t_\mathrm{int}=4.0\times 10^7 P_\ii$ here.

\begin{figure*}
 	\begin{center}
 	\includegraphics[clip,width=10cm]{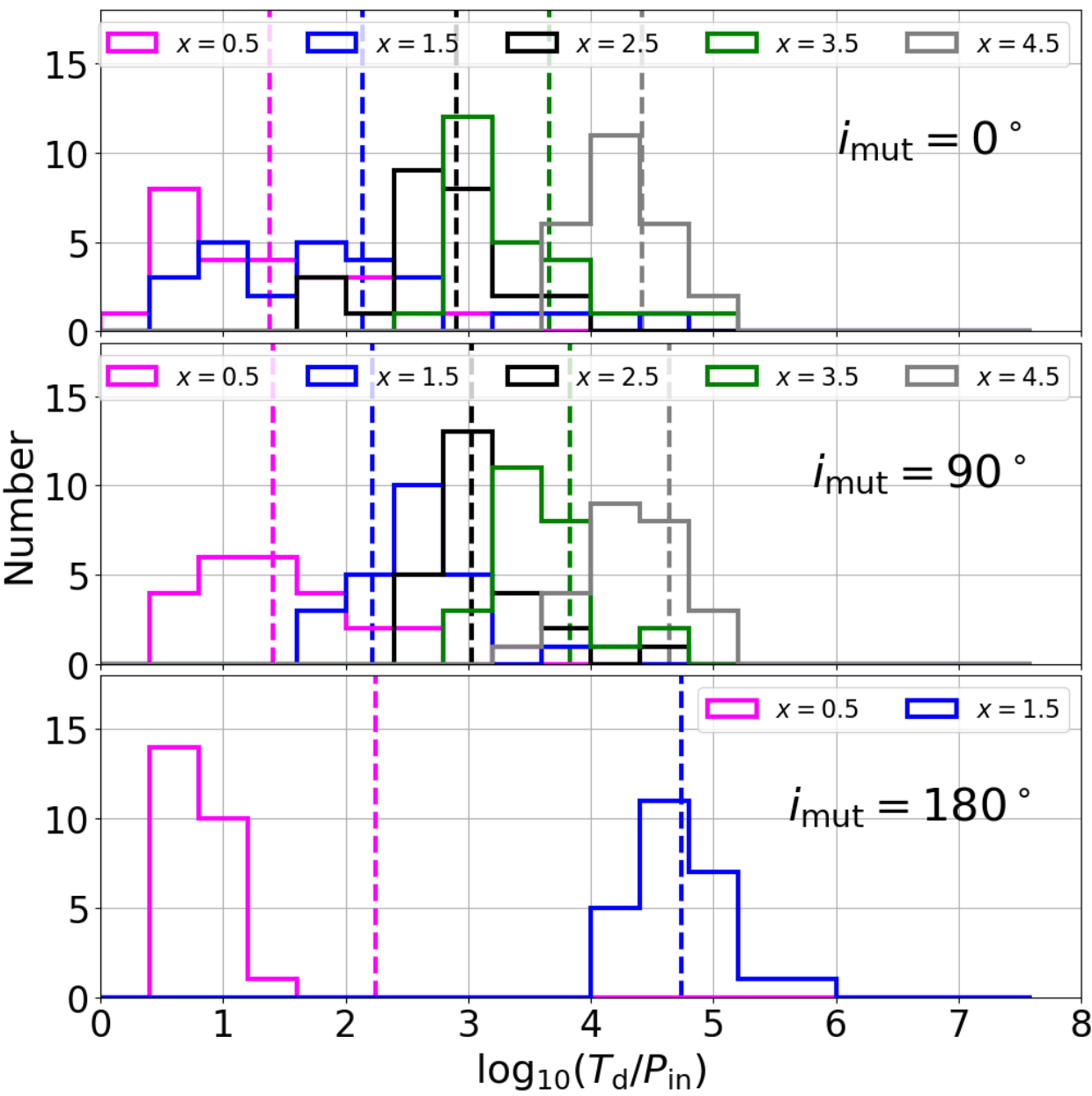}
 	\end{center}
\caption{Histograms of the disruption timescales for various initial
  phase angles ($M_\oo,\omega_\oo$), corresponding to coplanar
  prograde (top), orthogonal (middle), and coplanar retrograde
  (bottom) triples with $q_{21}=1$, $q_{23}=5$, and $e_\oo=0.7$. The
  dashed vertical lines are from our empirical best-fit models in
  Figure \ref{fig:distribution_y}.
 	\label{fig:cdf_phase}}
\end{figure*}

As expected, the initial phases change the disruption timescales for
unstable triples mostly within one or two order-of-magnitude. We also
performed similar runs for orthogonal and coplanar retrograde triples,
and plotted the histograms of $\log_{10}(T_\mathrm{d}/P_\ii)$ in
Figure \ref{fig:cdf_phase}; initially coplanar prograde (top),
orthogonal (middle), and coplanar retrograde (bottom). The figure
implies that our result in the main text is confirmed statistically,
within one or two order-of-magnitude scatters similar to those coming
from the chaotic behavior of those triples.

\begin{figure*}
 	\begin{center}
 	\includegraphics[clip,width=15cm]{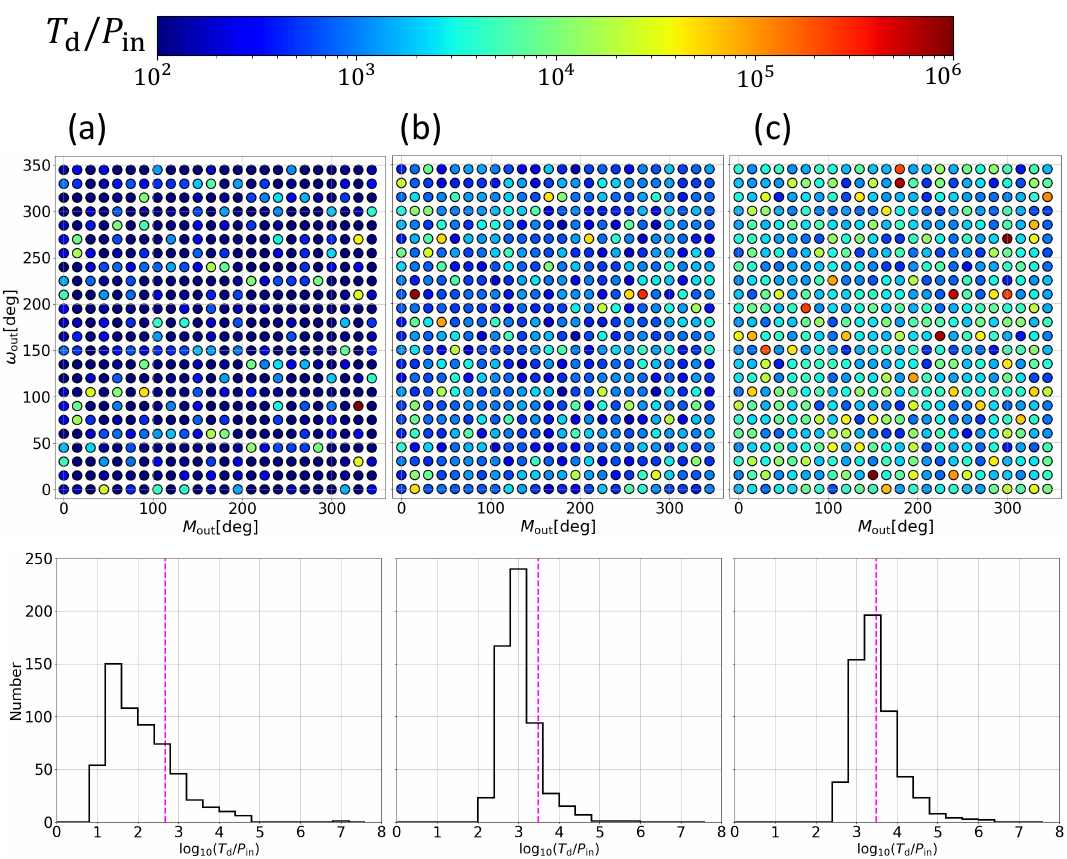}
 	\end{center}
\caption{The disruption timescale distributions of triples for
    various phase angles ($M_\oo,\omega_\oo$) (top), and the
    corresponding histograms (bottom). We adopt the same values for
    the other initial orbital parameters as the panels (a), (b) and
    (c) in Figure \ref{fig:digit}. The dashed vertical lines are from
    our empirical best-fit models in Figure
    \ref{fig:distribution_y}. Note that $T_\mathrm{d}/P_\ii<10^2$ and
    $T_\mathrm{d}/P_\ii>10^6$ are assigned the same color as $10^2$
    and $10^6$, respectively.
 	\label{fig:phase_pattern}}
\end{figure*}

Furthermore, we examine how the initial phases affect the the
disruption timescales. For that purpose, we consider the three
specific examples corresponding to panels (a), (b), and (c) in Figure
\ref{fig:digit} with $P_\ii=1000$ days. Then, we vary their initial
phases $(M_\oo,\omega_\oo)$ systematically between $0^\circ$ and
$360^\circ$ in grids of $15^\circ$. The resulting disruption
timescales are plotted in Figure \ref{fig:phase_pattern} on $M_\oo$ -
$\omega_\oo$ plane (top), together with the corresponding histograms
(bottom). Those plots show that the disruption timescales vary with
the initial phases, within one or two order-of magnitudes. The
sensitivity to the initial phase dependence does not seem to be
directly correlated to the chaotic behavior that is clearly shown in
Figure \ref{fig:digit}; the panel (c) exhibits the similar scatters of
the disruption timescale as the other two cases, while it does not
show any chaotic behavior against the tiny change in the initial
$P_\ii$.

\section{Ejection pattern of three bodies}\label{sec:ejection}

\begin{figure*}
\begin{center}
		\includegraphics[clip,width=8cm]{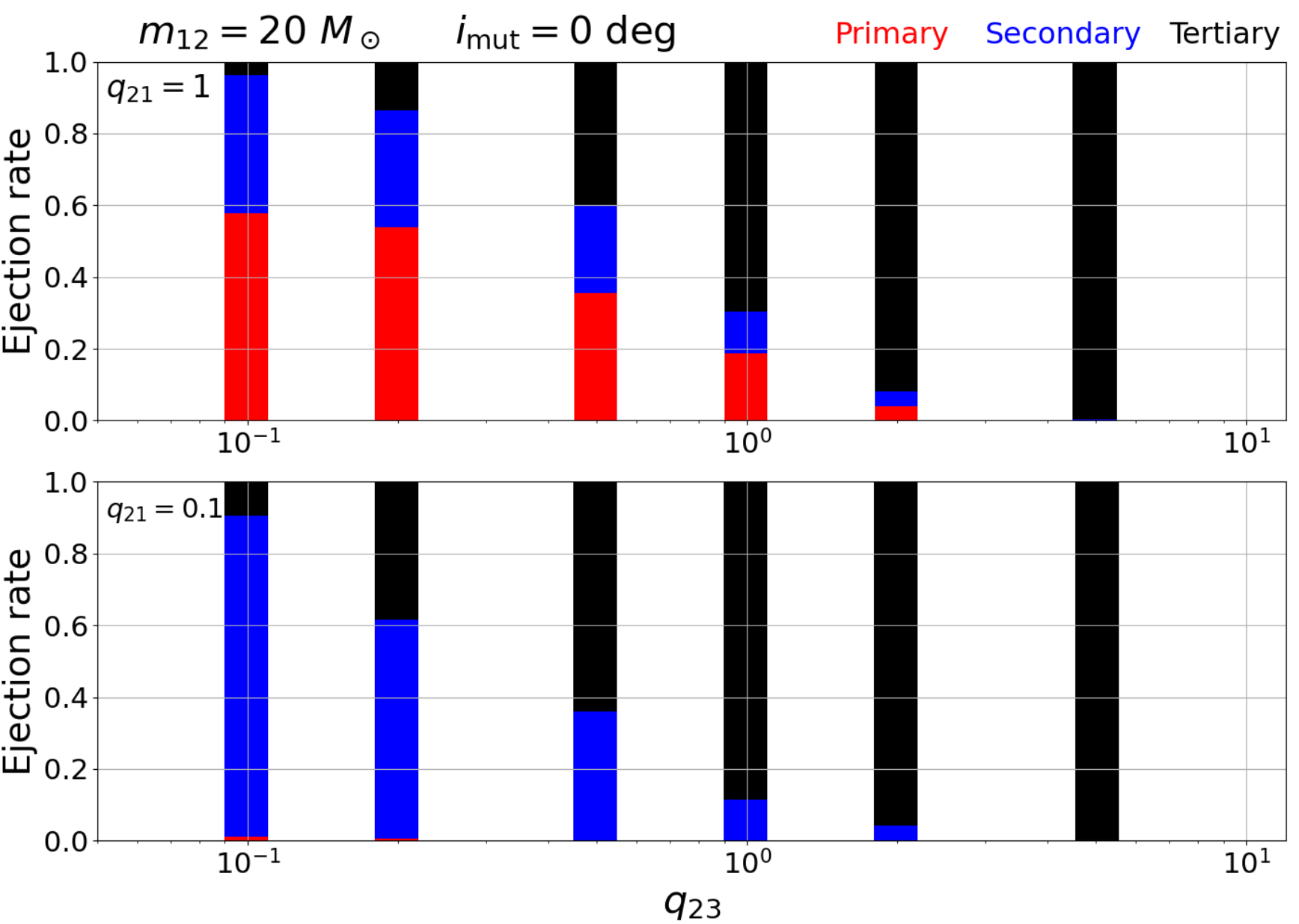}
		\hspace{100pt}
		\includegraphics[clip,width=8cm]{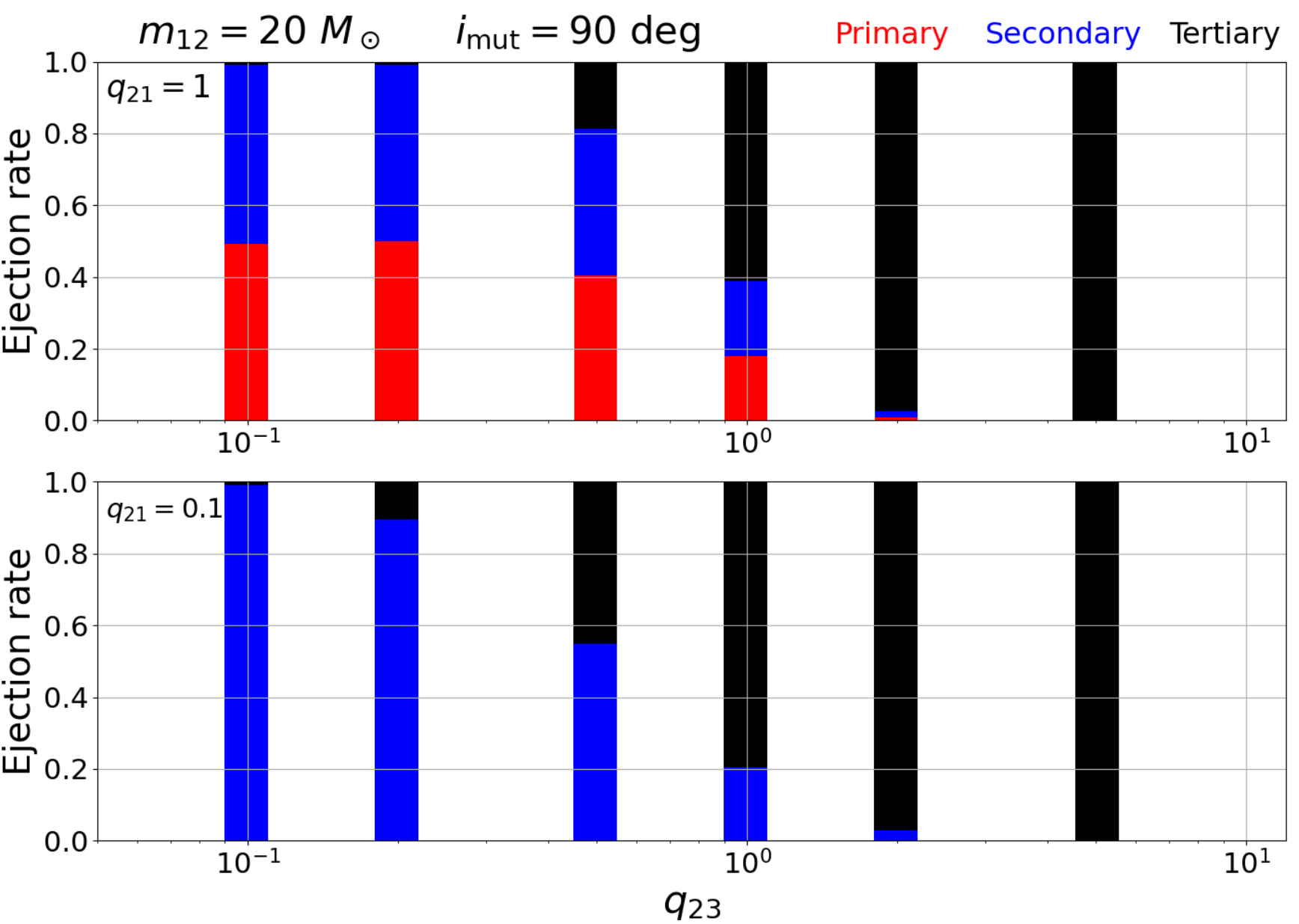}
		\hspace{100pt}
		\includegraphics[clip,width=8cm]{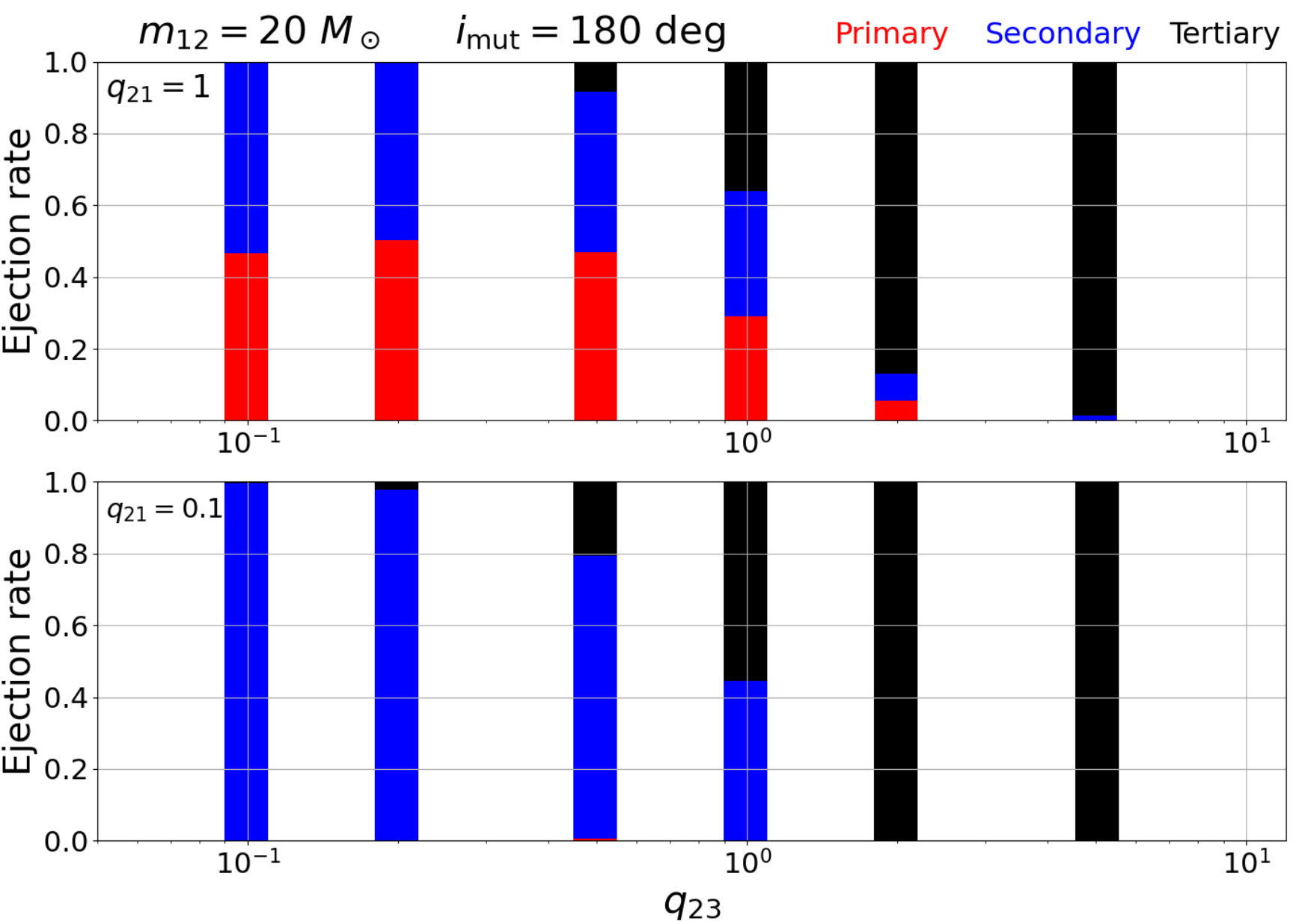}
\end{center}
\caption{Dependence of the ejection rate of three bodies on
    $q_{23}$. Red, blue, and black bars represent the ejection of the
    primary, secondary, and tertiary body, respectively. Top, middle,
    and bottom panels show prograde, orthogonal, and retrograde cases,
    respectively. In each set of panels, upper and lower plots
    correspond to $q_{21}=1.0$ and $0.1$, respectively.}
		\label{fig:ejection_rate}
\end{figure*}

Our result indicates that the disruption timescales and
  stability boundary are sensitive to the mutual inclinations, but
  very weakly on $q_{21}$ and $q_{23}$.  In order to examine the
  dependence on $q_{21}$ and $q_{23}$ from a different point of view,
  we show the fraction rate of bodies that are ejected from disrupted triples
  in Figure \ref{fig:ejection_rate}.

The primary ejection, secondary ejection, and tertiary
    ejection, are color-coded in red, blue, and black,
    respectively. In order to save the computational cost, we here
    adopt $t_\mathrm{int}=4.0\times10^7P_\ii$. Top, middle, and bottom
    panels show the results for prograde, orthogonal, and retrograde
    triples, respectively. Each panel contains the result for
    $q_{21}=1$ (upper) and $0.1$ (lower). We note that there is no
    physical reason to distinguish between the primary and secondary
    for $q_{21}=1$, and the small difference in their ejection rate
    should simply come from statistical reasons.

  As expected, the tertiary is preferentially ejected for as
    the value of $q_{23}$ increases.  The inner less massive body
    starts to dominate the ejection rate when $q_{23}< 0.5$. The
    dependence on $q_{23}$ is stronger for retrograde triples, which
    is is indeed consistent with Figures \ref{fig:distribution_y} and
    \ref{fig:fitting}.

\section{The timescales of GR precessions and GW emissions}
\label{sec:grprecession}

We discuss briefly the possible effects of general relativity (GR)
  that we neglect in the present analysis by considering the GR
  precession and gravitational-wave emission timescales.

It is known that both the quadrupole and octupole ZKL
  oscillation is strongly suppressed when the GR precession rate
  $\dot{\omega}_\mathrm{GR}$ becomes comparable to the ZKL
  precession rate $\dot{\omega}_\mathrm{ZKL}$. The ratio between the
  two precessions is explicitly given as \citep[e.g.][]{Liu2015}:
\begin{eqnarray}
\label{eq:precession-ratio}
  \frac{\dot{\omega}_\mathrm{GR}}{\dot{\omega}_\mathrm{ZKL}}&=&
  \frac{3(1-e_\oo^2)^{3/2}}{\sqrt{1-e_\ii^2}} \frac{r_\mathrm{s,\ii}}{2a_\ii}
  \frac{m_{12}}{m_3} \left(\frac{a_\ii}{a_\oo}\right)^{-3} \nonumber \\ 
  &\approx& 1.1\times 10^{-3} \left(\frac{a_\ii/a_\oo}{0.1}\right)^{-3}
  \left(\frac{2~M_\odot}{m_3}\right)\left(\frac{m_{12}}{20~M_\odot}\right)^2\left(\frac{a_\ii}{5.3~\mathrm{au}}\right)^{-1}
  \frac{(1-e_\oo^2)^{3/2}}{\sqrt{1-e_\ii^2}},
\end{eqnarray}
where $c$ is the speed of light, and
$r_\mathrm{s,\ii}\approx2.96 (m_{12}/M_\odot)~\mathrm{km}$ is the
Schwarzschild radius of inner binary as if it would be a single body. The estimate above was calculated using the initial conditions of our fiducial triple (see table~\ref{tab:fiducial}). Thus, it is not likely that GR
precession affects the evolution too much except for very massive or close inner binary.
Since GR precession suppresses the ZKL oscillation, preventing the oscillations in eccentricity that cause orthogonal triples to be more unstable (see figure~\ref{fig:A1A2A3}), GR precession may tend to stabilize the triples in general.

In addition, the long-term GW emission shrinks the orbit of the
  inner binary. Thus, it reduces the gravitational interaction between
  the inner and outer orbits, resulting in the stabilization of
  triples.  The typical GW emission timescale may be characterized by
  the following merger timescale \citep[][]{Peters1964} for a circular
  inner binary:
\begin{eqnarray}
  T_\mathrm{GW,0} &=& \frac{5}{32}
  \left(\frac{a_\ii}{r_\mathrm{s,\ii}}\right)^3
  \frac{m_{12}^2}{m_1m_2}\frac{a_\ii}{c} \nonumber \\
  &\approx& 3.17\times 10^{16} \frac{m_{12}^2}{m_1m_2}\left(\frac{m_{12}}{20~M_\odot}\right)^{-3}\left(\frac{a_\ii}{5.3~\mathrm{au}}\right)^4~\mathrm{yrs},
\end{eqnarray}
where $a_\ii$ is the initial inner semi-major axis, and we assume our fiducial triple of $a_\ii = 5.31$ au. Thus, it is not likely that GW emission affects the evolution for systems without the ZKL oscillation except for a very massive and close inner binary for circular case.
On the other hand, the eccentricity might be enhanced by the ZKL oscillations in highly inclined triples. The increase in eccentricity can strongly decrease the coalescence timescale of the inner binary. Given the maximum eccentricity $e_{\rm max}$ reached during a ZKL cycle, the merger timescale can be expressed as \citep[e.g.][]{Liu2018}:
\begin{eqnarray}
T_\mathrm{GW} = T_\mathrm{GW,0}(1-e_\mathrm{max}^2)^3,
\end{eqnarray}
If we set $e_\mathrm{max}=1-10^{-4}$, the merger
timescale decreases by eleven order-of-magnitudes. Thus, the
combination effect of the instability from the ZKL oscillation and
stabilization from GW emission would be very important to understand
the stability of triples for $i_\mathrm{mut}\sim 90^\circ$ and $q_{21}
\ll 1$.

Alternatively, the dynamical instability itself may trigger the enhancement of the eccentricity of the inner binary, or even lead to a temporary state in which no stable binary exist, but all three bodies interact gravitationally with each other. During such states, defined in the literature as democratic resonances \citep{Hut1983}, two BHs may obtain very close separation so that a merger ensues. This model of gravitational wave coalescence has been extensively studied in the context of three-body encounters in dense environments \citep[e.g.][]{Dicarlo2020}, but also in the context of destabilized field triples \citep[e.g.][]{Toonen2021}.


\end{document}